\documentclass[apj]{emulateapj-rtx4}

\shorttitle{}
\shortauthors{Kim et al.}
\begin{document}
\title{Medium-resolution Optical and Near-Infrared Spectral Atlas of\\
 16 2MASS-selected NIR-red Active Galactic Nuclei at $z \sim 0.3$}

 \author{	\textbf{Dohyeong Kim}\altaffilmark{1,2},		\textbf{Myungshin Im}\altaffilmark{1,2},
 			\textbf{Gabriela Canalizo}\altaffilmark{3},		\textbf{Minjin Kim}\altaffilmark{4},
 			\textbf{Ji Hoon Kim}\altaffilmark{5}, 			\textbf{Jong-Hak Woo}\altaffilmark{2},
 			\textbf{Yoon Chan Taak}\altaffilmark{1,2},		\textbf{Jae-Woo Kim}\altaffilmark{4},
 			and \textbf{Mariana Lazarova}\altaffilmark{6,7}}

\altaffiltext{1}{Center for the Exploration of the Origin of the Universe (CEOU),
Astronomy Program, Department of Physics and Astronomy, Seoul National University, 
1 Gwanak-ro, Gwanak-gu, Seoul 151-742, South Korea}
\altaffiltext{2}{Astronomy Program, Department of Physics and Astronomy, Seoul National University, 1 Gwanak-ro, Gwanak-gu, Seoul 151-742, South Korea}
\altaffiltext{3}{Department of Physics and Astronomy, University of California, Riverside, CA 92521, USA}
\altaffiltext{4}{Korea Astronomy and Space Science Institute, Daejeon 305-348, South Korea}
\altaffiltext{5}{Subaru Telescope, National Astronomical Observatory of Japan, 650 North A’ohoku Place, Hilo, HI 96720, USA}
\altaffiltext{6}{Department of Physics and Astronomy, University of Nebraska - Kearney, 2401 11th Avenue, Kearney, NE 68849, USA}
\altaffiltext{7}{Department of Physics and Astronomy, University of Northern Colorado, Greeley, CO 80639, USA}
\email{dohyeong@astro.snu.ac.kr; mim@astro.snu.ac.kr}

\begin{abstract}
 We present medium-resolution spectra ($R \sim 2000$--4000) at 0.4--1.0\,$\mu$m and 0.7--2.5\,$\mu$m of
 16 active galactic nuclei (AGNs) selected with red color in the near-infrared (NIR) of $J - K > 2.0$ mag at $z \sim 0.3$.
 We fit the H$\beta$, H$\alpha$, P$\beta$, and P$\alpha$ lines from these spectra
 to obtain their luminosities and line widths.
 We derive the $E(B-V)$ color excess values of the NIR-red AGNs using two methods,
 one based on the line-luminosity ratios and another based on the continuum slopes.
 The two $E(B-V)$ values agree with each other at rms dispersion $\sim$ 0.249.
 About half of the NIR-red AGNs have $g'-K < 5$ magnitude,
 and we find that these NIR red, but blue in optical-NIR AGNs, have $E(B-V) \sim 0$,
 suggesting that a significant fraction of the NIR color-selected red AGNs are unobscured or only mildly obscured.
 After correcting for the dust extinction, we estimate the black hole (BH) masses and
 the bolometric luminosities of the NIR-red AGNs using the Paschen lines to calculate their Eddington ratios ($\lambda_{\rm Edd}$).
 The median Eddington ratios of nine NIR-red AGNs ($\log (\lambda_{\rm Edd}) \simeq -0.654 \pm 0.176$)
 are only mildly higher than those of unobscured type 1 AGNs ($\log (\lambda_{\rm Edd}) \simeq -0.961 \pm 0.008$).
 Moreover, we find that the $M_{\rm BH}$--$\sigma_{\ast}$ relation for three NIR-red AGNs
 is consistent with that of unobscured type 1 AGNs at similar redshift.
 These results suggest that the NIR red color selection alone is not effective at picking up dusty, intermediate-stage AGNs.
\end{abstract}

\section{Introduction}
 In recent years, there has been increased interest in the role of supermassive black holes (SMBHs) in galaxy formation and evolution.
 Most massive spheroidal galaxies are believed to harbor SMBHs at their centers,
 and the masses of the SMBHs have a correlation with the luminosities \citep{magorrian98,bentz09,bennert10,greene10,jiang11,park15},
 the stellar velocity dispersions \citep{ferrarese00,gebhardt00,tremaine02,gultekin09,woo10},
 and Sersic indices \citep{graham01,graham07} of their spheroids.

 The proposed scenario for the growth of SMBHs is that they are assembled by gas accretion \citep{lynden69}.
 The SMBHs are thought to grow very rapidly in an active phase, i.e., the period of active galactic nuclei (AGNs).
 During this phase, AGNs emit enormous amounts of energy
 ($10^{43}$--$10^{48}$\,$\rm erg\,s^{-1}$; e.g., \citealt{woo02}) from gamma-ray to radio.
 Most of our knowledge of AGNs comes from unobscured type 1 AGNs,
 found by using surveys of X-ray, ultraviolet (UV), optical, and radio observations
 \citep{grazian00,becker01,anderson03,croom04,risaliti05,schneider05,veron-cetty06,young09}.

 However, several studies (e.g., \citealt{comastri01,tozzi06,polletta08}) have reported that
 the soft X-ray, UV, and optical based AGN surveys could neglect
 a large number (e.g., up to more than $50\%$) of AGNs with very red colors,
 due to the dust extinction from the intervening dust and gas in their host galaxies \citep{webster95,cutri02}.
 From a similar but different point of view,
 there is another missing population of AGNs with red colors due to the interstellar medium in our galaxy \citep{im07,lee08}.

\begin{figure*}[!t]
	\centering
	\figurenum{1}
	\includegraphics[width=\textwidth]{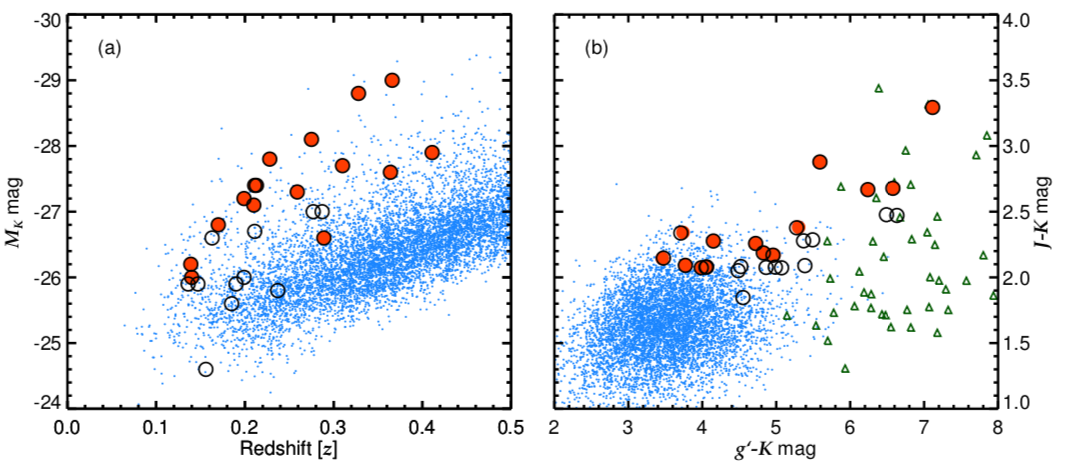}\\
	\caption{(a) Redshifts vs. $M_{K}$ magnitudes for NIR-red AGNs and unobscured type 1 quasars.
		The circles represent 29 NIR-red AGNs listed in \cite{marble03} and
		the red filled circles denote 16NIR-red AGNs used in this work.
		The blue dots represent the redshifts and $M_{K}$ magnitudes of unobscured type 1 quasars listed in SDSS DR7 \citep{shen11}.
		(b) Color-color diagram using $g'-K$ and $J-K$ magnitudes.
		The meanings of the circles and the blue dots are identical to the left panel,
		and the green triangles denote the $g'-K$ and $J-K$ magnitudes of red quasars used in our previous studies \citep{kim15b,kim18}.
		Here, $g'$-, $J$-, and $K$-band magnitudes are corrected for the Galactic extinction \citep{schlafly11},
		and the $g'$-band magnitude is in AB unit, while $J$- and $K$-band magnitudes are based on the Vega magnitudes.}
\end{figure*}

 The AGNs with red colors are called red AGNs, and they are considered to be a different population from unobscured type 1 AGNs.
 In several simulation studies \citep{hopkins05,hopkins06,hopkins08}, red AGNs have been predicted to be in an intermediate phase between
 the merger-driven star-forming galaxies, such as ultra-luminous infrared galaxies (ULIRGs; \citealt{sanders88,sanders96}),
 and unobscured type 1 AGNs.
 Although this explanation for red AGNs is still controversial due to several reasons
 (e.g., \citealt{puchnarewicz98,whiting01,wilkes02,schawinski11,schawinski12,simmons12,kocevski12,rose13}),
 this scenario is further supported by several observational studies.
 For example, red AGNs have (i) high accretion rates \citep{urrutia12,kim15b},
 (ii) enhanced star formation activity \citep{georgakakis09},
 (iii) a frequent occurrence of merging features \citep{urrutia08,glikman15},
 (iv) young radio jets \citep{georgakakis12},
 (v) red continua from dust extinction \citep{glikman07,urrutia09}, and
 (vi) line-luminosity ratios explainable only when considering dust extinction \citep{kim18}.

 Since the observational results of red AGNs could arise from the limited sample size of red AGNs,
 there have been several efforts to search for more red AGNs
 (\citealt{webster95,benn98,cutri01,cutri02,smith02,glikman07,glikman12,urrutia09,banerji12,stern12,assef13,fynbo13,lacy13}).
 A significant number of these studies found red AGNs using large-area NIR photometric surveys,
 such as the Two Micron All-Sky Survey (2MASS; \citealt{skrutskie06}),
 the UKIRT Infrared Deep Sky Survey (UKIDSS; \citealt{lawrence07}),
 and the $\it Wide$-$\it field~Infrared~Survey~Explorer$ ($\it WISE$; \citealt{wright10}) survey.

 Compared to how large-area NIR photometric surveys have contributed to our understanding of red AGNs,
 investigations based on NIR spectroscopic data
 (e.g., \citealt{glikman07,glikman12,kim15b}) are limited.
 Despite the limited contribution, the NIR spectrum includes useful information for investigating the nature of red AGNs.
 For example, (i) BH masses (Paschen lines: \citealt{kim10,landt11b}, and Brackett lines: \citealt{kim15a}),
 (ii) bolometric luminosities \citep{kim10},
 (iii) broad-line region (BLR) sizes \citep{landt11a},
 (iv) temperatures \citep{glikman06,landt11b,kim15a} and covering factors of hot dust \citep{kim15a},
 (v) stellar velocity dispersions ($\sigma_{\ast}$; \citealt{woo10,kang13}), and
 (vi) star forming activity \citep{imanishi11,kim12}
 can be measured from the NIR spectra.

 In this work, we present high signal-to-noise ratio (S/N; up to several hundreds) and medium-resolution ($R \sim 2000$)
 optical and NIR spectra of a sample of 16 NIR-red AGNs at $z \sim 0.3$,
 for which optical images and polarizations were obtained in previous studies \citep{smith02,marble03}.
 We concentrate on a detailed description of our sample and observation (\S\,2),
 spectral fittings for hydrogen lines (\S\,3), dust-reddening measurements (\S\,4),
 accretion rates (\S\,5), and the $M_{\rm BH}$--$\sigma_{\ast}$ relation  for NIR-red AGNs (\S\,6).
 In \S\,7, we briefly summarize our results.
 Throughout this work, we use a standard $\Lambda$CDM cosmological model of $H_{0}=70\,{\rm km\,s^{-1}}$ Mpc$^{-1}$,
 $\Omega_{m}=0.3$, and $\Omega_{\Lambda}=0.7$,
 supported by past observational studies (e.g., \citealt{im97}).
 Our photometry uses the Vega magnitude system, except for the $g'$ band that is in the AB system.

\section{The Sample and Observation}
\subsection{Sample}
 Our sample is drawn from the 29 2MASS-based red AGNs listed in \cite{marble03}.
 The 29 objects were selected with the following procedures.
 First, \cite{cutri01,cutri02} chose red AGN candidates through a combination of red color in NIR ($J-K_{\rm s} > 2$)
 and detection in each of the three 2MASS bands (complete to $K_{s} < 15.0$\,mag).
 Then, among the candidates, 70 targets were spectroscopically confirmed in \cite{smith02}.
 Furthermore, \cite{smith02} performed optical polarimetric observations
 using the Two-Holer Polarimeter/Photometer on the Steward Observatory 1.5\,m telescope and the Bok 2.3\,m reflector.
 Finally, \cite{marble03} selected 29 out of the 70 NIR-red AGNs within the redshift range of $0.136 \leq z \leq 0.596$,
 and observed them with the Wide Field Planetary Camera 2 (WFPC2) on board the $\it Hubble~Space~Telescope$ ($\it HST$).

\begin{deluxetable*}{ccccccccccccc}[!t]
	\tabletypesize{\scriptsize}
	\tablecolumns{10}
	\tablewidth{0pt}
	
	\tablenum{1}
	
	\tablecaption{Observing summary \label{tbl1}}
	\tablewidth{0pt}
	\tablehead{
		\colhead{}&	\colhead{}&	\colhead{}&	\colhead{}&	\colhead{}&	\colhead{}&	\multicolumn{3}{c}{NIR Spectroscopy}&	\colhead{}&	\multicolumn{3}{c}{Optical Spectroscopy}\\
		\cline{7-9} \cline{11-13}\\
		\colhead{Object}&		\colhead{R.A.}&			\colhead{Decl.}&
		\colhead{Redshift}&		\colhead{$K_{\rm s}$}&	\colhead{$M_{K_{\rm s}}$\tablenotemark{a}}&
		\colhead{Telescope/}&	\colhead{Exp}&			\colhead{Observing}&	\colhead{}&		\colhead{Telescope/}&	\colhead{Exp}&	\colhead{Observing}\\
		\colhead{}&				\colhead{(J2000.0)}&	\colhead{(J2000.0)}&
		\colhead{($z$)}&		\colhead{(mag)}&		\colhead{(mag)}&
		\colhead{Instrument}&	\colhead{(s)}&		\colhead{dates}&		\colhead{}&		\colhead{Instrument}&	\colhead{(s)}&	\colhead{dates}}
	
	\startdata
	0106$+$2603&	01 06 07.7&	+26 03 34&	0.411&	14.6&	-27.9&	Subaru/IRCS&	800\tablenotemark{b}&	2015 Nov&	&	Keck/ESI&	3600&	2003 Oct\\
	&				&			&		&		&		&	&				6000\tablenotemark{c}&	2015 Nov&	&	\\
	0157$+$1712&	01 57 21.0&	+17 12 48&	0.213&	13.2&	-27.4&	Gemini/GNIRS&	3600&	2015 Aug&	&	Keck/ESI&	7200&	2003 Oct\\
	0221$+$1327&	02 21 50.6&	+13 27 41&	0.140&	13.2&	-26.0&	Magellan/FIRE&	6657&	2015 Jan&	&	Keck/ESI&	5400&	2003 Oct\\
	0234$+$2438&	02 34 30.6&	+24 38 35&	0.310&	13.7&	-27.7&	Gemini/GNIRS&	2160&	2015 Aug&	&	--&			--&		--\\
	0324$+$1748&	03 24 58.2&	+17 48 49&	0.328&	12.8&	-28.8&	Magellan/FIRE&	3635&	2015 Jan&	&	Keck/ESI&	3600&	2004 Sep\\
	0348$+$1255&	03 48 57.6&	+12 55 47&	0.210&	13.6&	-27.1&	Gemini/GNIRS&	2880&	2015 Aug&	&	Keck/ESI&	3600&	2004 Sep\\
	1258$+$2329&	12 58 07.4&	+23 29 21&	0.259&	13.4&	-27.3&	IRTF/SpeX&		9000&	2016 Mar&	&	SDSS\\
	1307$+$2338&	13 07 00.6& +23 38 05&	0.275&	13.4&	-28.1&	IRTF/SpeX&		18000&	2016 Mar&	&	--&			--&		--\\
	1453$+$1353&	14 53 31.5&	+13 53 58&	0.139&	13.1&	-26.2&	IRTF/SpeX&		9600&	2016 Mar&	&	SDSS\\
	1543$+$1937&	15 43 07.7&	+19 37 51&	0.228&	12.7&	-27.8&	Gemini/GNIRS&	3600&	2016 Apr&	&	Keck/ESI&	3600&	2004 Jul\\
	&				&			&		&		&		&	IRTF/SepX&		9600&	2016 Mar&	&	\\
	1659$+$1834&	16 59 39.7&	+18 34 36&	0.170&	12.9&	-26.8&	IRTF/SpeX&		8400&	2016 Mar&	&	Keck/ESI&	5400&	2004 Jul\\
	2222$+$1952&	22 22 02.2&	+19 52 31&	0.366&	13.3&	-29.0&	Gemini/GNIRS&	2160&	2015 Aug&	&	--&			--&		--\\
	2222$+$1959&	22 22 21.1&	+19 59 47&	0.211&	12.9&	-27.4&	Gemini/GNIRS&	2880&	2015 Aug&	&	Keck/ESI&	5400&	2004 Sep\\
	2303$+$1624&	23 03 04.3&	+16 24 40&	0.289&	14.7&	-26.6&	Gemini/GNIRS&	5040&	2015 Aug&	&	--&			--&		--\\
	2327$+$1624&	23 27 45.6&	+16 24 34&	0.364&	14.5&	-27.6&	Gemini/GNIRS&	3600&	2015 Aug&	&	Keck/ESI&	5400&	2004 Sep\\
	2344$+$1221&	23 44 49.5&	+12 21 43&	0.199&	12.9&	-27.2&	Gemini/GNIRS&	2880&	2015 Aug&	&	Keck/ESI&	3600&	2004 Jul
	\enddata
	\tablenotetext{a}{The $M_{K_{\rm s}}$ values are recalculated using the method of Marble et al. (2003)
		with the standard $\Lambda$CDM cosmological model}
	\tablenotetext{b}{The grism mode observation}
	\tablenotetext{c}{The echelle mode observation}
\end{deluxetable*}

 Among the 29 NIR-red AGNs, we select 16 AGNs at $z \sim 0.3$ (from 0.139 to 0.411)
 for which the redshifted P$\beta$ or P$\alpha$ line is observable within the sky window wavelength range.
 The 16 NIR-red AGNs span over a wide range of luminosities ($-29.0 < M_K < -26.0$). 
 Figure 1 shows the redshifts versus the $M_{K}$ magnitudes and $g'-K$ colors versus the $J-K$ colors of
 the 16 NIR-red AGNs and red quasars used in our previous studies (\citealt{kim15b,kim18}; originally from \citealt{urrutia09}).
 These NIR-red AGNs have red colors of $J-K > 2$ and $g'-K \gtrsim 4$.
 Compared to the red quasars that we studied previously ($J-K > 1.3$ and $g'-K > 5$),
 a non-negligible fraction of the 16 NIR-red AGNs possess $g'-K < 5$ as blue as those of unobscured type 1 quasars.

 For our sample, we emphasize the advantage of the availability of various types of high-quality data.
 The 16 NIR-red AGNs have optical images from $\it HST$ \citep{marble03} and
 optical broadband polarimetry \citep{smith02}.
 We expect that a combined data set of the high-quality images from the $\it HST$ data, the optical polarization,
 and the optical/NIR spectra from this study
 will be unique and useful for the comprehensive investigations of the nature of NIR-red AGNs.

\subsection{NIR Observation}
 We performed NIR spectroscopic observations with four telescopes and their respective instruments.
 Since the 16 NIR-red AGNs have different brightnesses and redshifts,
 the observations need to be performed with proper observational instruments and telescopes to fit the characteristics of the 16 NIR-red AGNs.
 We describe the details of our NIR observations below.

 First, NIR spectra of nine NIR-red AGNs were obtained with the cross-dispersed mode of
 the Gemini Near-infrared Spectrograph (GNIRS; \citealt{elias06}) on the 8.1\,m Gemini-North telescope.
 The observational configuration is a combination of a 110 l/mm grating, short blue camera, and 0$\farcs$675 slit width,
 which provides a discontinuous spectral coverage from $\sim 1$\,$\mu$m to $\sim 2.1$\,$\mu$m
 with a spectral resolution of $R \sim 2600$.

 Second, we used the SpeX \citep{rayner03} on the 3.0\,m NASA Infrared Telescope Facility (IRTF) for five NIR-red AGNs.
 In this observation, we used the short cross-dispersion mode (SXD) with a 0$\farcs$3 slit width
 to achieve a spectral resolution of $R \sim 2000$ across 0.7\,--\,2.55\,$\mu$m.
 Among these five NIR-red AGNs, one, 1543$+$1937, overlaps with the nine NIR-red AGNs observed with GNIRS/Gemini.

 Third, in order to obtain the NIR spectra of two NIR-red AGNs,
 we used the Folded-port Infrared Echellette (FIRE) on the 6.5\,m Magellan Baade telescope with a 1$\farcs$0 slit width.
 This observational configuration allows the wavelength coverage to span from 0.82 to 2.51\,$\mu$m
 with a resolving power of $R \sim 3600$.

 Fourth, an NIR spectrum of one NIR-red AGN was obtained with
 the Infrared Camera and Spectrograph (IRCS; \citealt{tokunaga98,kobayashi00}) on the 8.2\,m Subaru telescope.
 In this observation, we used both grism and echelle modes.
 For the grism mode observation, we used a 0$\farcs$1 slit width and $HK$ band with the grism of 52 milliarcsecond pixel scale,
 which provides a spectral coverage of 1.4\,--\,2.5\,$\mu$m with a spectral resolution of $R \sim 440$.
 The Echelle mode observation was performed with a 0$\farcs$54 slit width and $K$ band,
 and this provides a spectral resolution of $R \sim 6600$ with a discontinuous wavelength coverage from 1.90\,$\mu$m to 2.49\,$\mu$m.

 \begin{figure*}
	\centering
	\figurenum{2}
	\includegraphics[width=\textwidth]{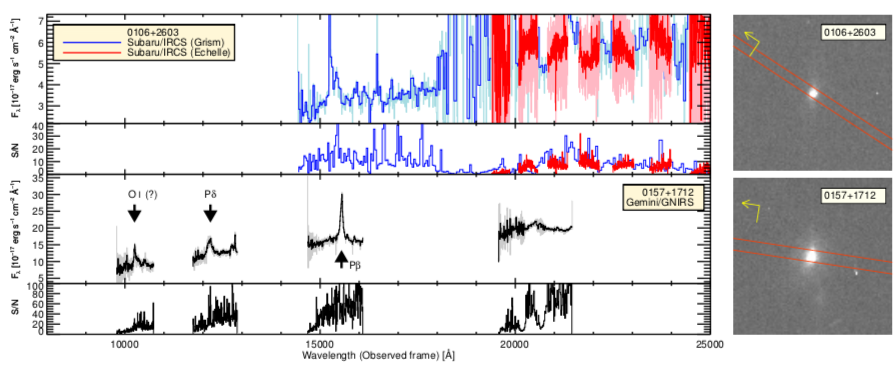}\\
	\includegraphics[width=\textwidth]{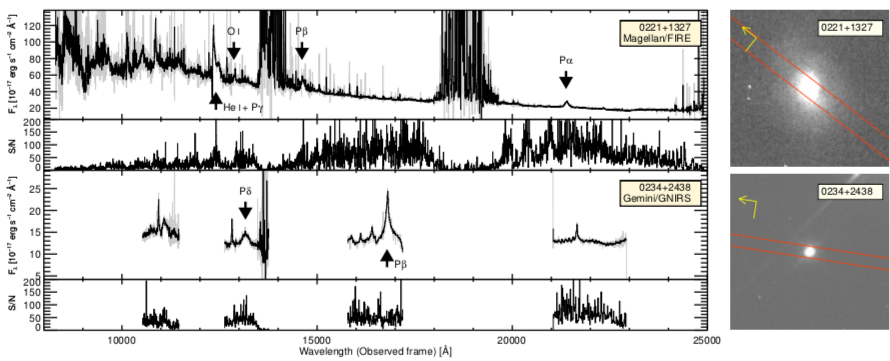}\\
	\includegraphics[width=\textwidth]{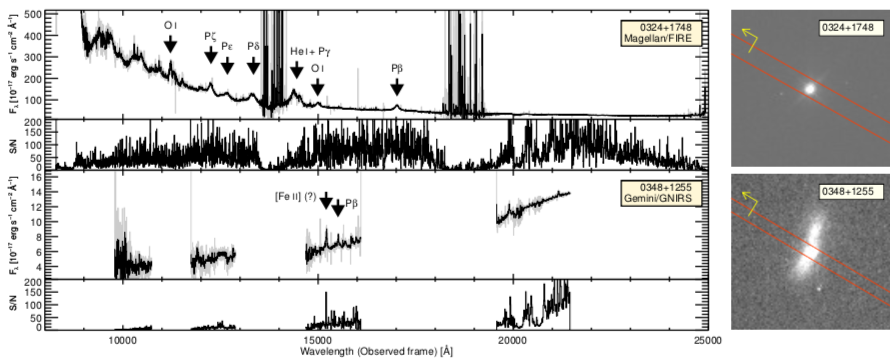}\\
	\caption{(Left) NIR spectra of 16 NIR-red AGNs and their S/Ns. The gray lines indicate observed spectra in the observed frame,
		and the black lines indicate binned spectra with the spectral resolution.
		For 0106$+$2603 and 1543$+$1937, the spectra were obtained with two different observing modes and instruments.
		Each binned spectrum from an individual observation is represented by the blue and red lines,
		and the sky blue and pink lines indicate their original spectra, respectively.
		Moreover, several emission lines (P$\alpha$: 1.8751\,$\mu$m, P$\beta$: 1.2818\,$\mu$m, P$\gamma$: 1.0938\,$\mu$m, P$\delta$: 1.0049\,$\mu$m,
		P$\epsilon$: 0.9546\,$\mu$m, P$\zeta$: 0.9229\,$\mu$m, [\ion{Fe}{2}]: 1.5995 and 1.2567\,$\mu$m,
		\ion{O}{1}: 1.1287, and 0.8446\,$\mu$m, and \ion{He}{1}: 1.0830\,$\mu$m) are marked on the spectra.
		However, when the emission line is not obvious due to the low S/N or duplicate sky lines,
		the emission line is marked with a question mark.
		(Right) $HST$ images of 16 NIR-red AGNs.
		The red boxes across the objects indicate slit widths and the yellow arrows at the top left denote north.}
\end{figure*}

\begin{figure*}[!t]
	\centering
	\figurenum{2}
	\includegraphics[width=\textwidth]{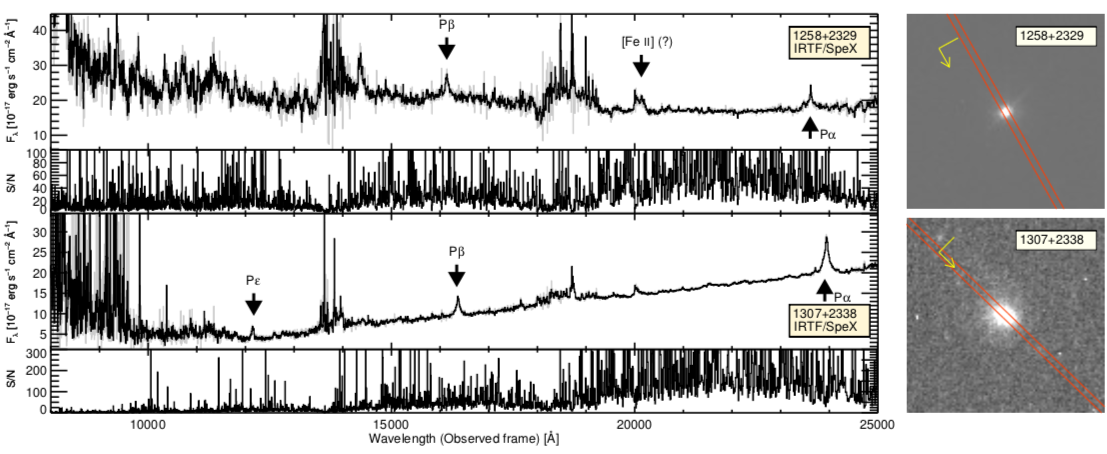}\\
	\includegraphics[width=\textwidth]{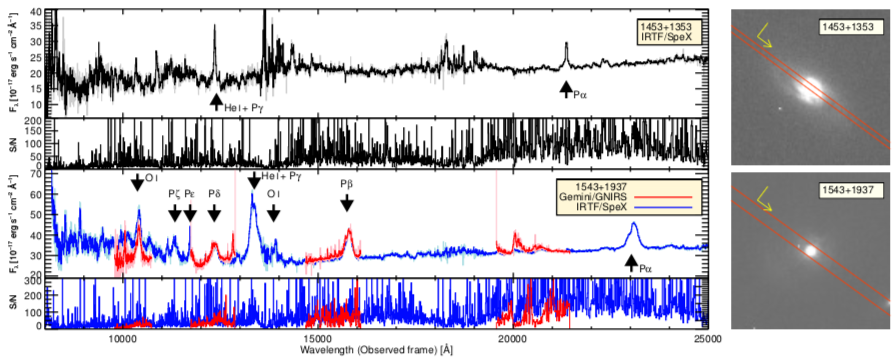}\\
	\includegraphics[width=\textwidth]{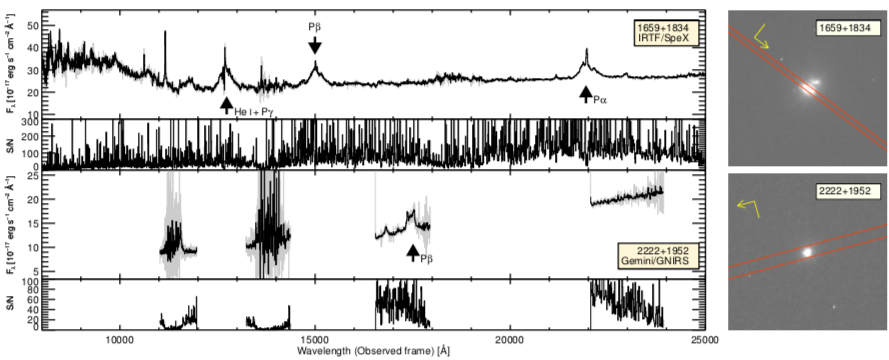}\\
	\caption{Continued}
\end{figure*}

\begin{figure*}[!t]
	\centering
	\figurenum{2}
	\includegraphics[width=\textwidth]{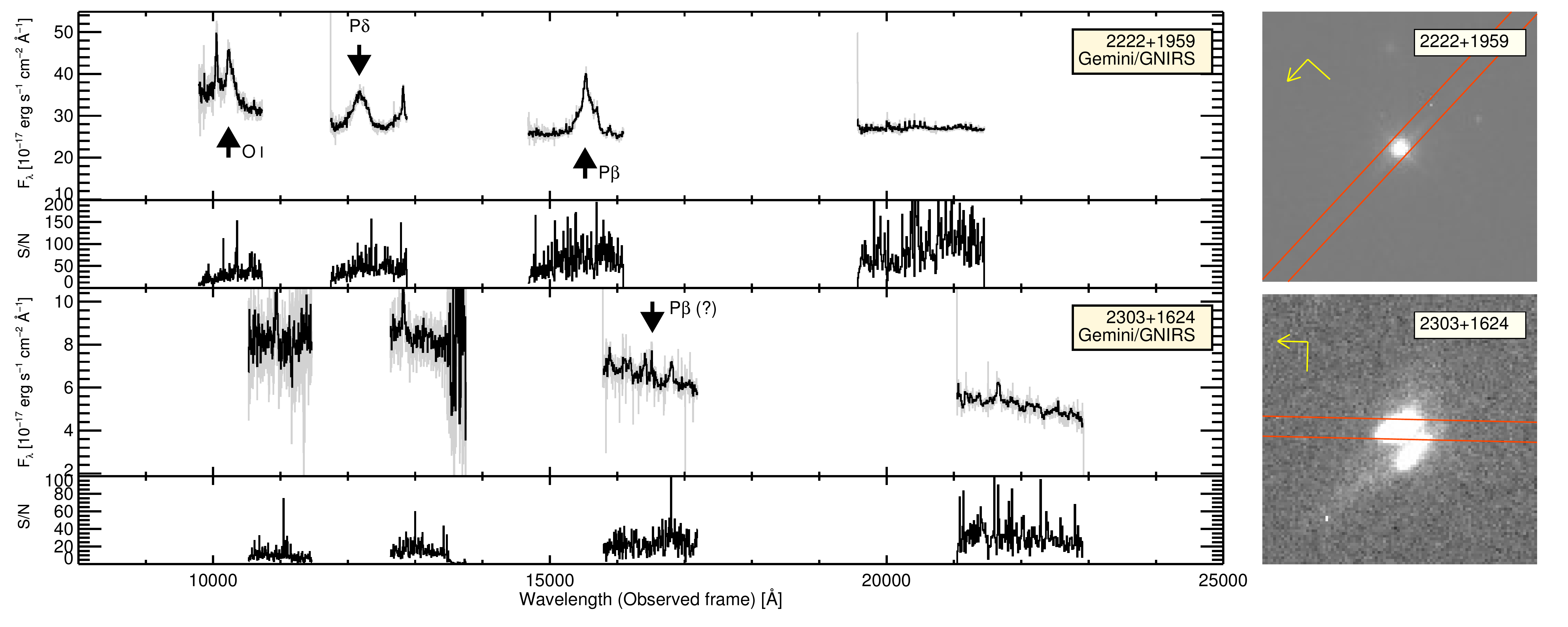}\\
	\includegraphics[width=\textwidth]{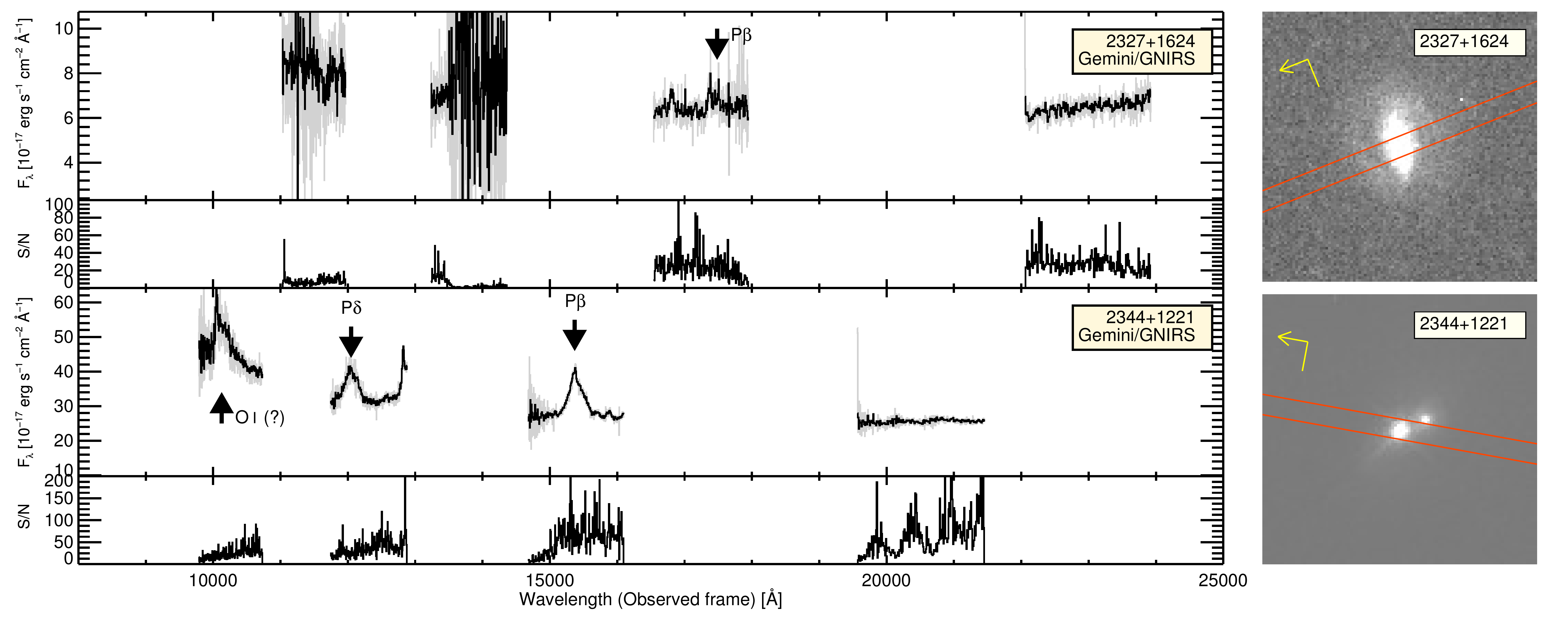}\\
	\caption{Continued}
\end{figure*}

 Our observations were performed under clear weather conditions with sub-arcsecond seeings of $\sim$$0 \farcs$6.
 For the flux calibration and telluric correction, we observed nearby A0V stars before or after the observations of the NIR-red AGNs.
 In order to produce fully reduced spectra, we used Gemini image reduction and analysis facility packages \citep{cooke05}, Spextool \citep{vacca03,cushing04},
 FIREHOSE, and general reduction procedure of spectra with image reduction and analysis facility \citep{massey88}
 for the spectra obtained with GNIRS, SpeX, FIRE, and IRCS, respectively.

  Note that the Gemini reduction packages do not include a flux calibration process for the GNIRS data,
 and it recommends achieving the flux calibration by scaling the observed spectrum to existing photometry or spectrum.
 For this reason, we perform the flux calibration by scaling the GNIRS spectrum to
 its $J$-, $H$-, and $K$-band magnitudes of 2MASS photometry.
 For 2303$+$1624, however, the $J$-band flux was abnormally small in comparison to
 the $H$- and $K$-band fluxes and had a large error,
 so we used $y$-band flux from the Pan-STARRS survey \citep{chambers16} instead of the $J$-band flux.
 Due to our observational setup, the GNIRS spectra were taken at four disjointed orders,
 and each order was flux-scaled using an adjacent band.
 The two short orders ($\sim$10,000--14,000\,$\rm \AA{}$) were scaled by $J$-band magnitude,
 and we used $H$- and $K$-band magnitudes for the third-order ($\sim$15,000--18,000\,$\rm \AA{}$)
 and fourth-order ($\sim$20,000--24,000\,$\rm \AA{}$), respectively.
 In order to check the reliability of the above GNIRS flux calibration,
 we obtained the SpeX spectrum for 1543+1937 (a bright and point-like AGN in our data, see the $HST$ image in Figure 2).
 Note that both SpeX and GNIRS data were obtained under clear weather and a decent seeing condition ($\sim 0 \farcs 7$).
 The SpeX data were flux-scaled using a single relative scaling factor using its $K$-band magnitude (see the paragraph below).
 Our comparison shows that the SpeX and GNIRS data match with each other within an $\sim 3$\,\% difference from
 the second through the fourth orders.
 However, the shortest order of the GNIRS spectrum was $\sim$20\,\% different from the SpeX data.
 Although the shortest order of GNIRS spectra were not used in this study,
 we caution that the flux calibration of the shortest order could be off by about 20\,\%.

 In the next step, the NIR spectra from SpeX, FIRE, and IRCS
 were scaled to their $K$-band magnitudes of 2MASS photometry.
 This step was necessary, since there is a possibility of photon loss due to
 the different observing conditions and slit widths.
 We find that the scaling factors for the SpeX, FIRE, and IRCS spectra are not significant,
 with factors of 0.91 to 1.74 (a median of 1.18),
 and in particular, the scaling factors of the spectra obtained with SpeX are somewhat smaller with $\sim$1.07 (from 0.91 to 1.32).
 We also note that the AGN variability in NIR is a worrisome factor in this kind of calibration,
 since we are calibrating the spectra using the data that were taken at a different epoch.
 However, the NIR AGN variability is known to be generally small ($\sim 0.2$\,mag; e.g., \citealt{enya02}),
 so this flux calibration should be good to $\sim$20\,\% accuracy.

\begin{deluxetable}{ccc}
	\tablecolumns{3}
	\tablewidth{0pt}
	\tablenum{2}
	\tablecaption{NIR Spectrum of 0106$+$2603\label{tbl2}}
	\tablehead{
		\colhead{$\lambda$}&		\colhead{$f_{\lambda}$}&							\colhead{$f_{\lambda}$ Uncertainty}\\
		\colhead{($\rm \AA{}$)}&	\colhead{($\rm erg\,s^{-1}\,cm^{-2}\,\AA{}$)}	&	\colhead{($\rm erg\,s^{-1}\,cm^{-2}\,\AA{}$)}
	}
	
	\startdata
	14439&	4.6018E-17&	2.2044E-17\\
	14471&	2.8404E-17&	4.8711E-18\\
	14504&	1.9541E-17&	4.8643E-18\\
	14537&	3.7478E-17&	6.2486E-18\\
	14570&	2.5344E-17&	3.7455E-18\\
	14603&	2.6977E-17&	1.8525E-18\\
	14636&	3.1782E-17&	2.4565E-18
	\enddata
	\tablecomments{This table represents only a part of the NIR spectrum of 0106$+$2603.
		All the NIR spectra of the 16 NIR-red AGNs obtained with the four telescopes
		are available in machine-readable format.}
\end{deluxetable}

 In total, we obtained 0.7--2.55\,$\mu$m NIR spectra of 16 NIR-red AGNs at $z \sim 0.3$
 with a moderate resolution of $R >$ 2000 from the four instruments and telescopes.
 We summarize the observation information in Table 1.

\subsection{Optical Observation}
In addition to the NIR spectra, we obtained optical spectra for 12 NIR-red AGNs.
Two (G. Canalizo \& M. Lazarova) of us observed 10 NIR-red AGNs
using the Echellette Spectrograph and Imager (ESI; \citealt{sheinis02}) on the Keck II telescope
with a spectral wavelength range of 3900\,$\rm \AA{}$ to 11000\,$\rm \AA{}$ and
a slit width of 1$\farcs$0 to achieve a spectral resolution of $R \sim 4000$.
Descriptions of the observations for the 10 NIR-red AGNs are given in Table 1.
Information about the data reduction is given in \citet{canalizo12}.
Among the 10 spectra from the Keck/ESI observation, 5 were used in \citet{canalizo12}.

For the remaining two NIR-red AGNs, we obtained the optical spectra from Data Release 12 (DR12) of the Sloan Digital Sky Survey (SDSS).
The fiber diameter is 3$\farcs$0, and
the spectral coverage of the SDSS spectra is 3800\,$\rm \AA{}$ to 9200\,$\rm \AA{}$ with a spectral resolution of 1500--2500.

 \begin{figure*}
	\centering
	\figurenum{3}
	\includegraphics[width=\textwidth]{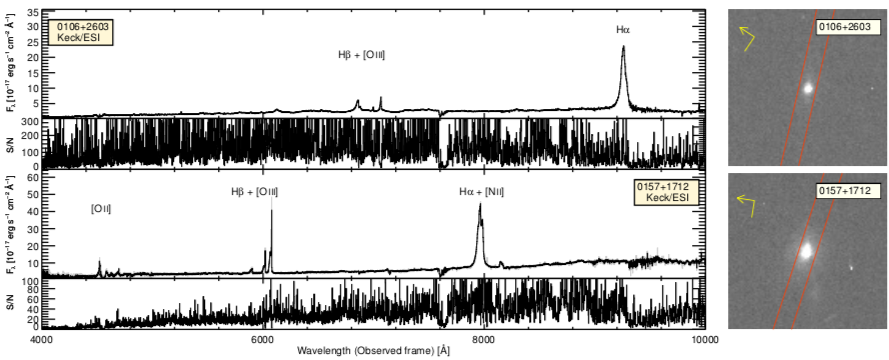}\\
	\includegraphics[width=\textwidth]{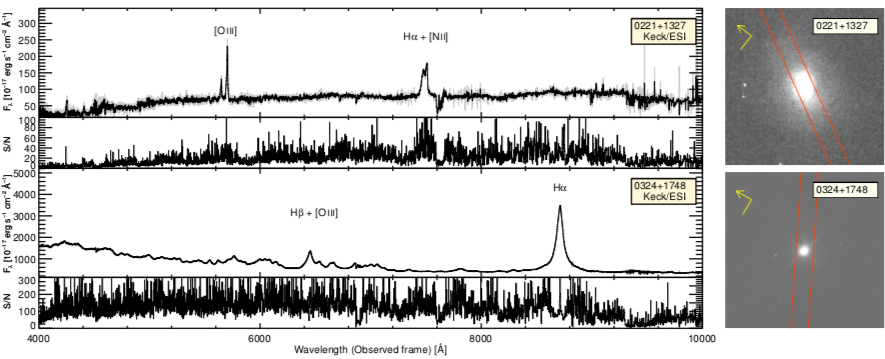}\\
	\includegraphics[width=\textwidth]{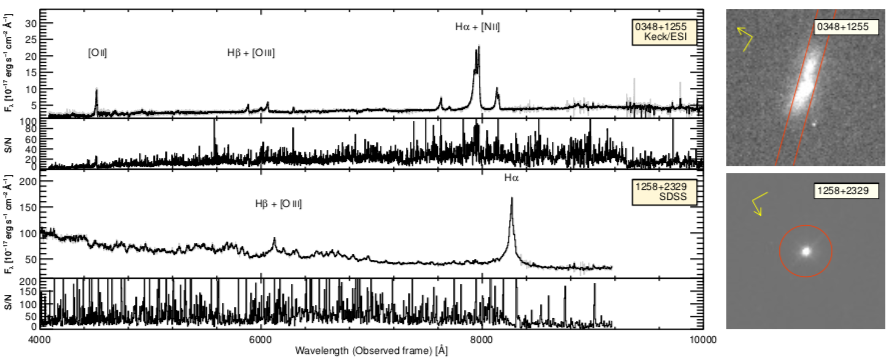}\\
	\caption{(Left) Optical spectra of 12 NIR-red AGNs and their S/N.
		The gray and black lines indicate observed and binned spectra in the observed frame, and we binned the spectra with the spectral resolution.
		We mark several emission lines in the optical wavelength region,
		such as [\ion{O}{2}] (3727\,$\rm \AA{}$), H$\gamma$ (4340\,$\rm \AA{}$), H$\beta$ (4861\,$\rm \AA{}$),
		[\ion{O}{3}] (4959 and 5007\,$\rm \AA{}$), H$\alpha$ (6563\,$\rm \AA{}$), and [\ion{N}{2}] (6548 and 6583\,$\rm \AA{}$).
		(Right) $HST$ images of 12 NIR-red AGNs.
		The red boxes and yellow arrows are identical to those in Figure 2,
		and the red open circles indicate SDSS fiber diameters.}
\end{figure*}

\begin{figure*}
	\centering
	\figurenum{3}
	\includegraphics[width=\textwidth]{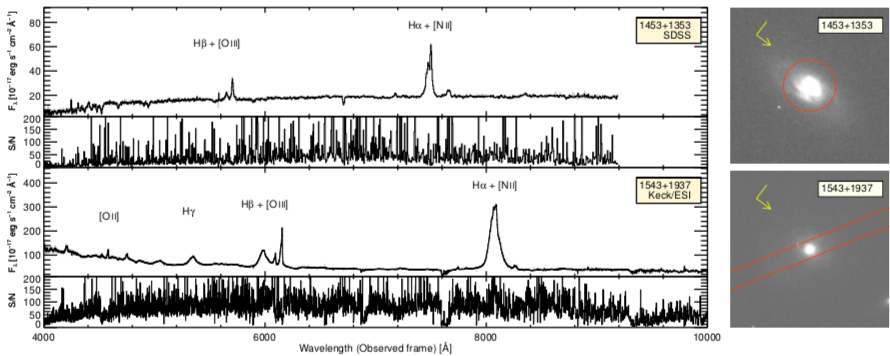}\\
	\includegraphics[width=\textwidth]{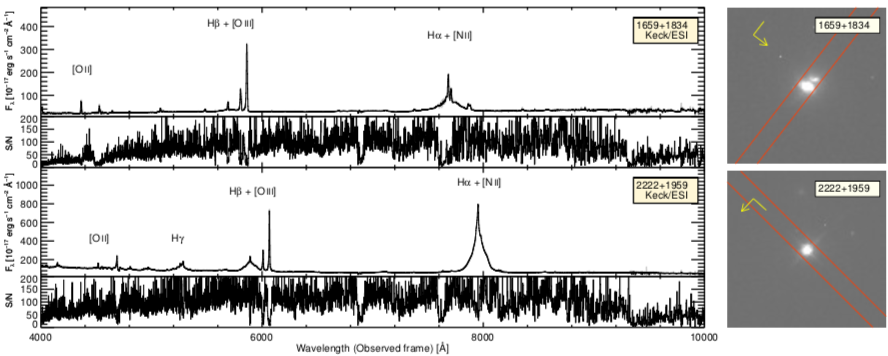}\\
	\includegraphics[width=\textwidth]{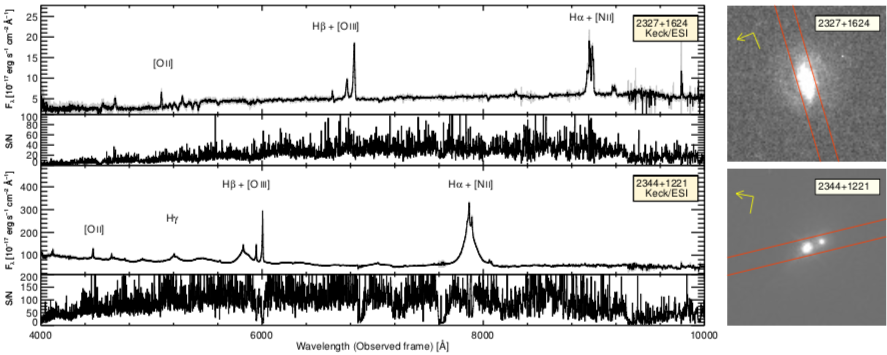}\\
	\caption{Continued}
\end{figure*}

For these optical spectra, their slit widths and fiber sizes are somewhat larger than
the slit widths for the NIR spectra.
Although the inconsistency of the slit width can introduce discrepancies in the spectral properties between the optical and NIR spectra,
the effects on the spectral properties measured in this study are negligible due to the following reasons:
(i) broad emission lines (BELs) come from the nuclear region, and
(ii) we use only the optical or NIR spectrum to fit the continuum.

\section{High-S/N and Medium-resolution Spectra}
 We show the fully reduced NIR spectra and the $HST$ images of the 16 NIR-red AGNs in Figure 2,
 and the spectra are available in machine-readable form in  Table 2.
 Moreover, Figure 2 also shows the S/N of each spectrum, and we mark several interesting lines
 such as P$\alpha$ (1.875\,$\mu$m), P$\beta$ (1.282\,$\mu$m), P$\gamma$ (1.094\,$\mu$m), P$\delta$ (1.005\,$\mu$m),
 P$\epsilon$ (0.955\,$\mu$m), P$\zeta$ (0.923\,$\mu$m), [\ion{Fe}{2}] (1.600 and 1.257\,$\mu$m),
 \ion{O}{1} (1.129 and 0.845\,$\mu$m), and \ion{He}{1} (1.083\,$\mu$m).

 In addition to the NIR spectra, Figure 3 shows the reduced optical spectra of the 12 NIR-red AGNs,
 and several lines of [\ion{O}{2}] (3727\,$\rm \AA{}$), H$\gamma$ (4340\,$\rm \AA{}$), H$\beta$ (4861\,$\rm \AA{}$),
 [\ion{O}{3}] (4959 and 5007\,$\rm \AA{}$), H$\alpha$ (6563\,$\rm \AA{}$), and [\ion{N}{2}] (6548 and 6583\,$\rm \AA{}$)
 are marked on the spectra. The optical spectra are given in ascii format  in Table 3.

\begin{figure*}[!t]
	\centering
	\figurenum{4}
	\includegraphics[scale=0.25]{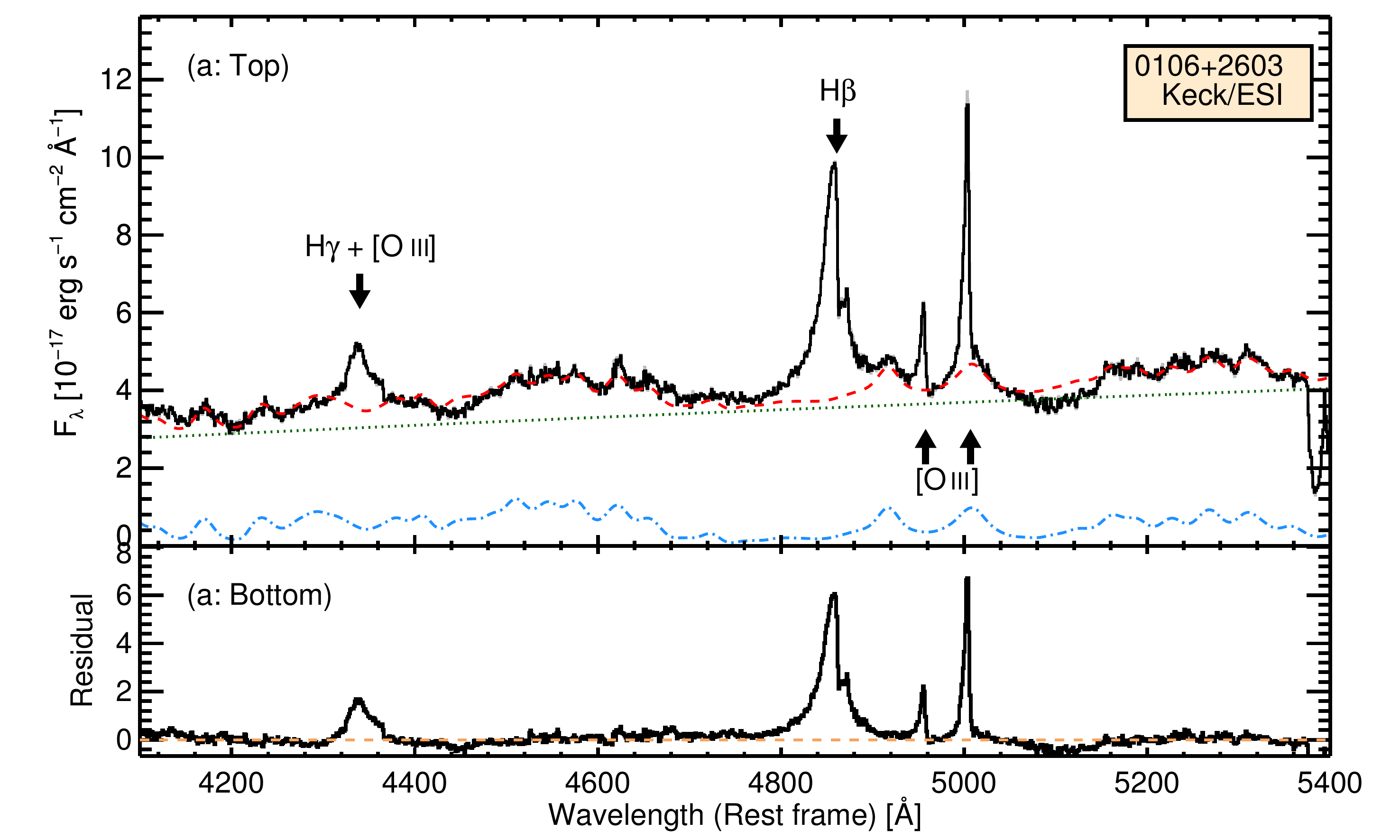}
	\includegraphics[scale=0.25]{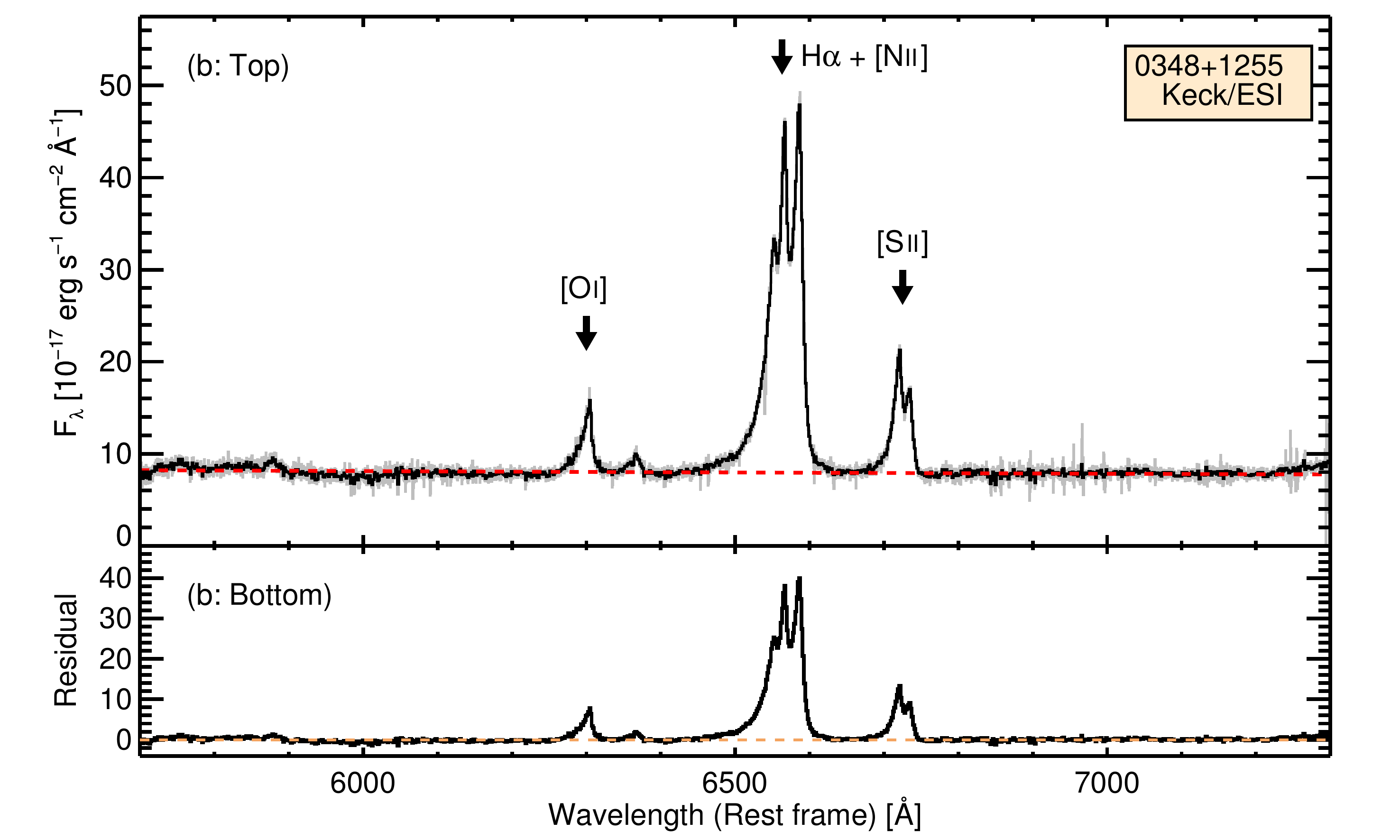}\\
	\includegraphics[scale=0.25]{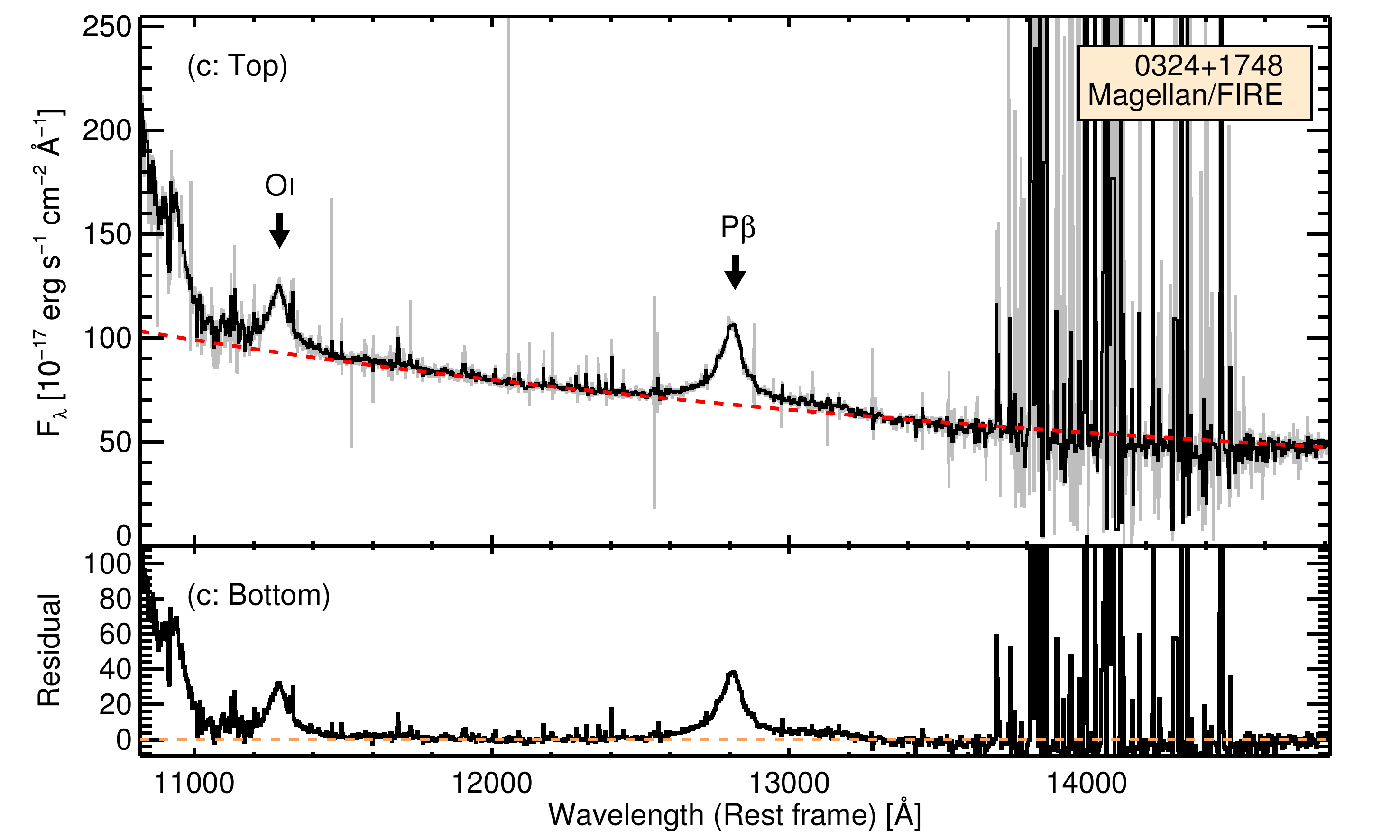}
	\includegraphics[scale=0.25]{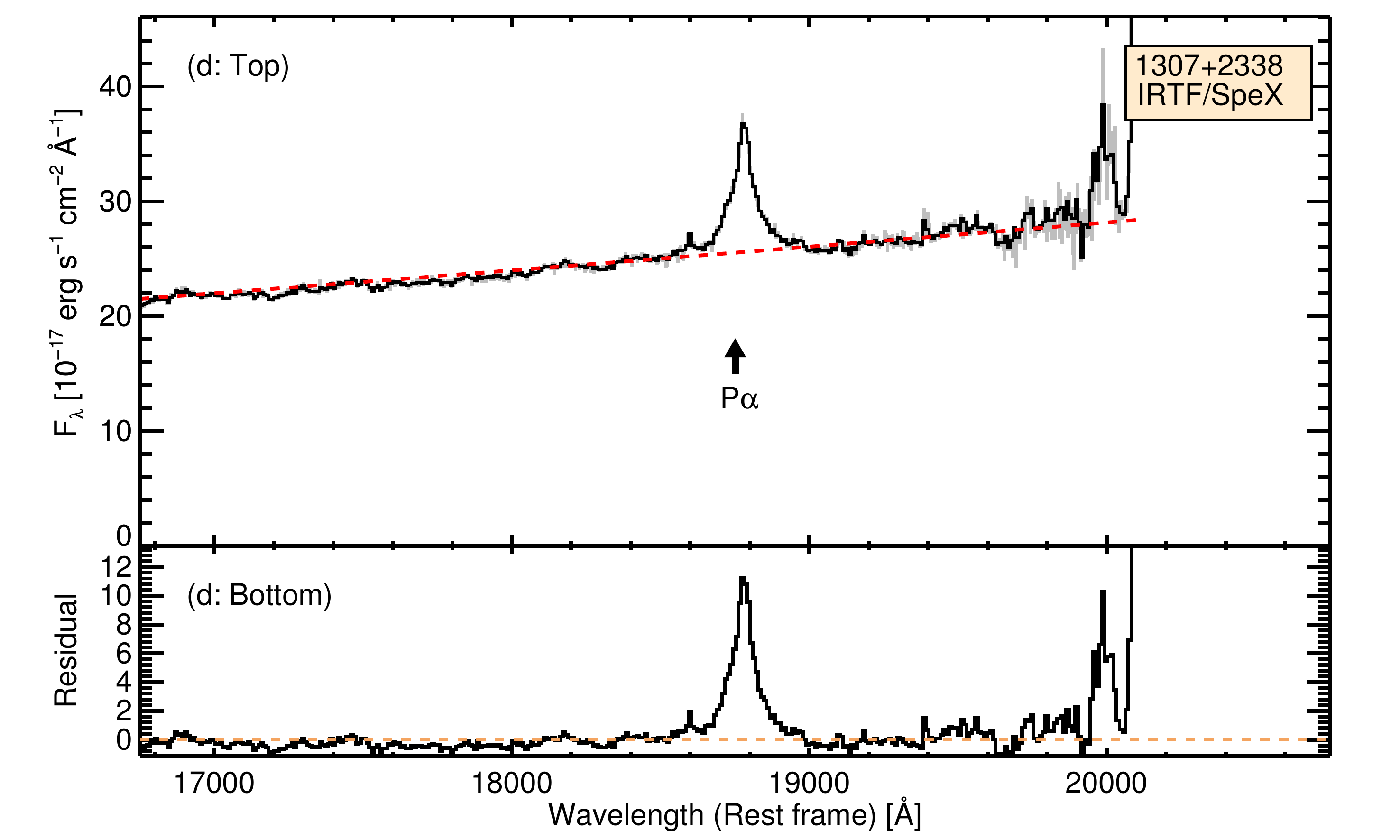}
	\caption{ 
		(a) Top: the optical spectrum of 0106$+$2603 obtained with the ESI on Keck telescope.
		The black solid line denotes the optical spectrum around the H$\beta$ line, and this spectrum includes several interesting emission lines,
		such as H$\gamma$ (4341\,$\rm \AA{}$), [\ion{O}{3}] (4363, 4959, and 5007\,$\rm \AA{}$), and H$\beta$ (4861\,$\rm \AA{}$).
		The fitted model continuum spectrum is represented by the red dashed line,
		and this model is composed of a power-law component (the green dotted line)
		and a component for the Fe blends (the sky blue dot-dashed line).
		Bottom: the fitted model continuum-subtracted spectrum is represented,
		and this continuum-subtracted spectrum is used for estimating the line luminosity and FWHM.
		(b) The original and residual spectrum around the H$\alpha$ line of 0348$+$1255, which is obtained with Keck/ESI.
		The meanings of the black solid line and the red dashed line are identical to those in panel (a).
		The black arrows represent [\ion{O}{1}] (6300\,$\rm \AA{}$), H$\alpha$ (6563\,$\rm \AA{}$),
		[\ion{N}{2}] (6548 and 6583\,$\rm \AA{}$), and [\ion{S}{2}] (6716 and 6731\,$\rm \AA{}$).
		(c) The original and residual spectrum around the P$\beta$ line of 0324$+$1748, which is taken with Magellan/FIRE.
		The meanings of the black solid line and the red dashed line are identical to those in panel (a),
		and P$\beta$ and \ion{O}{1} (1.1287\,$\mu$m) are marked with the black arrows.
		(d) The original and residual spectrum around the P$\alpha$ line of 1307$+$2338, which is obtained with IRTF/SpeX.
		The meanings of the black solid line and the red dashed line are identical to those in panel (a),
		and the P$\alpha$ line is marked with the black arrow.
	}
\end{figure*}

\begin{deluxetable}{ccc}
	\tablecolumns{3}
	\tablewidth{0pt}
	\tablenum{3}
	\tablecaption{Optical Spectrum of 0106$+$2603\label{tbl3}}
	\tablehead{
		\colhead{$\lambda$}&		\colhead{$f_{\lambda}$}&							\colhead{$f_{\lambda}$ Uncertainty}\\
		\colhead{($\rm \AA{}$)}&	\colhead{($\rm erg\,s^{-1}\,cm^{-2}\,\AA{}$)}	&	\colhead{($\rm erg\,s^{-1}\,cm^{-2}\,\AA{}$)}
	}
	
	\startdata
	3899.4&	8.3145E-17&	6.4623E-18\\
	3900.4&	5.3173E-17&	1.1044E-17\\
	3901.3&	2.6999E-17&	3.4251E-18\\
	3902.3&	6.7212E-17&	1.8250E-17\\
	3903.3&	1.2357E-16&	1.4382E-17\\
	3904.3&	1.4238E-16&	3.2512E-18\\
	3905.2&	1.1617E-16&	1.4047E-17 
	\enddata
	\tablecomments{This table represents only a part of the optical spectrum of 0106$+$2603.
		The optical spectra of the 12 NIR-red AGNs from the Keck/ESI observation and SDSS data
		are available in machine-readable format.}
\end{deluxetable}

\subsection{Spectral Fitting of Hydrogen Lines}
 In this subsection, we describe how the BELs of H$\beta$, H$\alpha$, P$\beta$, and P$\alpha$ are fitted
 to measure the luminosities and FWHMs.
 The fitting of these lines starts with the identification of the line,
 and we find the H$\beta$, H$\alpha$, P$\beta$, and P$\alpha$ lines in 11, 12, 14, and 6 NIR-red AGNs, respectively. 

 After the line identification, we correct the spectra for the Galactic extinction \citep{schlafly11}
 using the reddening law of \cite{fitzpatrick99}.
 Then, we transform the spectra to the rest-frame and
 fit the continua for the H$\beta$, H$\alpha$, P$\beta$, and P$\alpha$ lines.
 For H$\alpha$, P$\beta$, and P$\alpha$,
 the continuum around each line is fitted with a single power law,
 however, an additional Fe component is required for the H$\beta$ line.
 The Fe blends are determined by scaling and broadening the Fe template from the spectrum of IZw1 \citep{boroson92},
 and this procedure is performed with \texttt{MPFIT} \citep{markwardt09} using Interactive Data Language (IDL).
 As an example, Figure 4 shows the spectra around the H$\beta$, H$\alpha$, P$\beta$, and P$\alpha$ lines,
 along with the fitted continuum models, and the continuum-subtracted spectra.
 Note that we omit the stellar component for the continuum fit,
 since the stellar component can be fit by the power law in such narrow wavelength ranges.

 We note that several lines exist around the H$\beta$, H$\alpha$, P$\beta$, and P$\alpha$ lines
 (e.g., H$\gamma$ $\lambda$4340, [\ion{O}{3}] $\lambda\lambda$4956, 5007 doublet,
 [\ion{S}{2}] $\lambda\lambda$6716, 6731 doublet, and \ion{O}{1} $\lambda$11287).
 Hence, the continuum-fitting regions are chosen to avoid the nearby lines.


 After the continuum subtraction, we model the narrow lines
 using the [\ion{O}{3}] $\lambda\lambda$4956, 5007 and [\ion{S}{2}] $\lambda\lambda$6716, 6731 doublets as templates.
 In order to fit the [\ion{S}{2}] lines, we use two single Gaussian functions.
 However, the [\ion{O}{3}] lines require double Gaussian functions for their asymmetric blue wings \citep{greene05}.
 Although \cite{greene05} suggest that the [\ion{O}{3}] lines are not appropriate as a template for
 the narrow lines of unobscured type 1 quasars due to the blue wings,
 this template gives better results
 than the [\ion{S}{2}] template when fitting the narrow components of the H$\beta$ line,
 for a part of our NIR-red AGN sample.
 However, since the [\ion{S}{2}] template fits the narrow components of H$\alpha$, P$\beta$, and P$\alpha$ lines better,
 the template from the [\ion{S}{2}] lines is primarily used to fit these lines,
 except when the [\ion{S}{2}] lines are not detected.

\begin{deluxetable*}{ccccccccccccc}[!t]
	\tablecolumns{13}
	\tablewidth{0pt}
	\tablenum{4}
	\tablecaption{Line measurements of [\ion{O}{3}], [\ion{N}{2}], and [\ion{S}{2}] lines \label{tbl4}}
	\tablehead{
		\colhead{Object Name}&	\colhead{}&
		\multicolumn{2}{c}{[\ion{O}{3}] $\lambda$5007}&	\colhead{}&
		\multicolumn{2}{c}{[\ion{N}{2}] $\lambda$6548}&	\colhead{}&
		\multicolumn{2}{c}{[\ion{S}{2}] $\lambda$6716}\\
		\cline{3-4} \cline{6-7} \cline{9-10}\\
		\colhead{}&	\colhead{}&	
		\colhead{$L$}&	\colhead{FWHM}&	\colhead{}&
		\colhead{$L$}&	\colhead{FWHM}&	\colhead{}&
		\colhead{$L$}&	\colhead{FWHM}\\
		\colhead{}&	\colhead{}&
		\colhead{($\rm{10^{38}\,erg\,s^{-1}}$)}&	\colhead{($\rm km\,s^{-1}$)}&	\colhead{}&
		\colhead{($\rm{10^{38}\,erg\,s^{-1}}$)}&	\colhead{($\rm km\,s^{-1}$)}&	\colhead{}&
		\colhead{($\rm{10^{38}\,erg\,s^{-1}}$)}&	\colhead{($\rm km\,s^{-1}$)}
	}
	
	\startdata
	0106$+$2603&	&	29.72$\pm$1.45& 	359.9$\pm$36.2&	&	11.75$\pm$0.11&		359.9$\pm$36.2&	&	--&						--				\\
	0157$+$1712&	&	41.72$\pm$4.01&		988.4$\pm$63.8&	&	11.58$\pm$0.14&		563.7$\pm$20.7&	&	7.510$\pm$0.395&		563.7$\pm$20.7	\\
	0221$+$1327&	&	154.6$\pm$12.3&		539.5$\pm$54.9&	&	22.73$\pm$0.46&		539.5$\pm$54.9&	&	--&						--				\\
	0234$+$2438&	&	--&					--&				&	--&					--&				&	--&					--					\\
	0324$+$1748&	&	--&					--&				&	--&					--&				&	--&					--					\\
	0348$+$1255&	&	21.54$\pm$0.74&		718.7$\pm$25.7&	&	20.81$\pm$0.20&		595.8$\pm$6.8&	&	20.62$\pm$0.35&		595.8$\pm$6.8		\\
	1258$+$2329&	&	--&					--&				&	--&					--&				&	--&					--					\\
	1307$+$2338&	&	--&					--&				&	--&					--&				&	--&					--					\\
	1453$+$1353&	&	--&					--&				&	10.25$\pm$0.38&		695.5$\pm$47.5&	&	4.699$\pm$0.349&	695.5$\pm$47.5		\\
	1543$+$1937&	&	483.2$\pm$23.5&		808.2$\pm$79.0&	&	55.58$\pm$0.80&		387.2$\pm$10.8&	&	19.70$\pm$1.03&		387.2$\pm$10.8		\\
	1659$+$1834&	&	344.2$\pm$5.4&		628.7$\pm$22.1&	&	30.97$\pm$0.35&		501.2$\pm$5.1&	&	27.22$\pm$0.40&		501.2$\pm$5.1		\\
	2222$+$1952&	&	--&					--&				&	--&					--&				&	--&					--					\\
	2222$+$1959&	&	987.2$\pm$41.4&		538.9$\pm$41.8&	&	84.91$\pm$0.84&		538.9$\pm$41.8&	&	--&					--					\\
	2303$+$1624&	&	--&					--&				&	--&					--&				&	--&					--					\\
	2327$+$1624&	&	116.3$\pm$20.8&		719.6$\pm$77.7&	&	29.92$\pm$0.52&		518.9$\pm$33.1&	&	13.76$\pm$1.59&		518.9$\pm$33.1		\\
	2344$+$1221&	&	248.9$\pm$10.1&		449.3$\pm$30.9&	&	52.69$\pm$0.99&		301.1$\pm$10.0&	&	17.00$\pm$0.97&		301.1$\pm$10.0		
	\enddata
	\tablecomments{The listed fluxes are not corrected from the dust extinction caused by their host galaxies.}
\end{deluxetable*}

 Moreover, the [\ion{S}{2}] narrow line template is also used for
 fitting the [\ion{N}{2}] $\lambda\lambda$6548, 6583 doublet and the narrow component of the hydrogen lines.
 For the fitting of the [\ion{N}{2}] lines,
 we fit the H$\alpha$ and [\ion{N}{2}] lines simultaneously.
 The width of the [\ion{N}{2}] line is fixed to the width of the narrow line template,
 and its flux ratio is fixed to 2.96 \citep{kim06}.

 We note that the narrow line template from the [\ion{O}{3}] lines is used for
 0106$+$2603 (H$\beta$ and H$\alpha$), 0157$+$1712 (H$\beta$), 0221$+$1327 (H$\beta$, H$\alpha$, P$\beta$, and P$\alpha$),
 0348$+$1255 (H$\beta$), 1659$+$1834 (H$\beta$), 2222$+$1959 (H$\beta$, H$\alpha$, and P$\beta$),
 2327$+$1624 (H$\beta$), and 2344$+$1221 (H$\beta$),
 and the narrow line template from the [\ion{S}{2}] lines is used for
 0157$+$1712 (H$\alpha$ and P$\beta$), 0348$+$1255 (H$\alpha$ and P$\beta$), 1453$+$1353 (H$\alpha$ and P$\alpha$),
 1543$+$1937 (H$\alpha$), 1659$+$1834 (H$\alpha$, P$\beta$, and P$\alpha$),
 2327$+$1624 (H$\alpha$), and 2344$+$1221 (H$\alpha$ and P$\beta$).
 The measured FWHMs and luminosities of the [\ion{O}{3}], [\ion{N}{2}], and [\ion{S}{2}] lines
 are listed in Table 4.

 For 1258$+$2329 (P$\alpha$) and 2303$+$1624 (P$\beta$),
 the narrow lines cannot be modeled due to the absence of the [\ion{O}{3}] and [\ion{S}{2}] lines,
 so Gaussian functions were fit to the lines.
 One of the fitted components is classified as the narrow component due to its FWHM being less than 600\,$\rm km\,s^{-1}$.

 Using the model of the narrow component, we simultaneously fit the broad-line
 ($\rm FWHM > 600\,km\,s^{-1}$) with a single or double Gaussian function.
 Figure 5 shows the H$\beta$, H$\alpha$, P$\beta$, and P$\alpha$ lines of NIR-red AGNs and its fitted models.
 For the fit, a single, double, or multiple Gaussian functions
 are used depending on the S/N and the resolution of the spectra.
 For example, many of the broad lines in NIR spectra are fitted with a single or double Gaussian function due to the limited spectral resolution.
 \cite{kim10} showed using 26 unobscured type 1 AGNs with high-S/N and high-resolution spectra that
 the application of the single/double component Gaussian fits to data with a limited spectral resolution and S/N
 can produce slightly biased line flux/FWHM values
 with respect to the multi-component ($> 2$ components) fits, which is the method to use
 if high-S/N, high-resolution data were available.
 Based on the results, they derived the correction factors to correct for the systematic bias,
 which are $\rm flux_{multi}$/$\rm flux_{double}=1.05$, $\rm flux_{multi}$/$\rm flux_{single}=1.06$,
 $\rm FWHM_{multi}$/$\rm FWHM_{double}=0.85$, and $\rm FWHM_{multi}$/$\rm FWHM_{single}=0.91$ \citep{kim10,kim18}.
 We adopt these values to convert the single/double Gaussian fit results to the multi-component fitting results.
 Moreover, the FWHMs are corrected for the instrumental resolution
 as $\rm FWHM^2=FWHM^2_{obs} - FWHM^2_{inst}$.

 We note that the FWHM of the P$\beta$ line of 2327$+$1624 is not measured,
 because the P$\beta$ line is fitted by two Gaussian components that are broadly split.
 The broadly split components yield four half-maximum points, so the FWHM cannot be measured.

\begin{figure*}
	\centering
	\figurenum{5}
	\includegraphics[width=\textwidth]{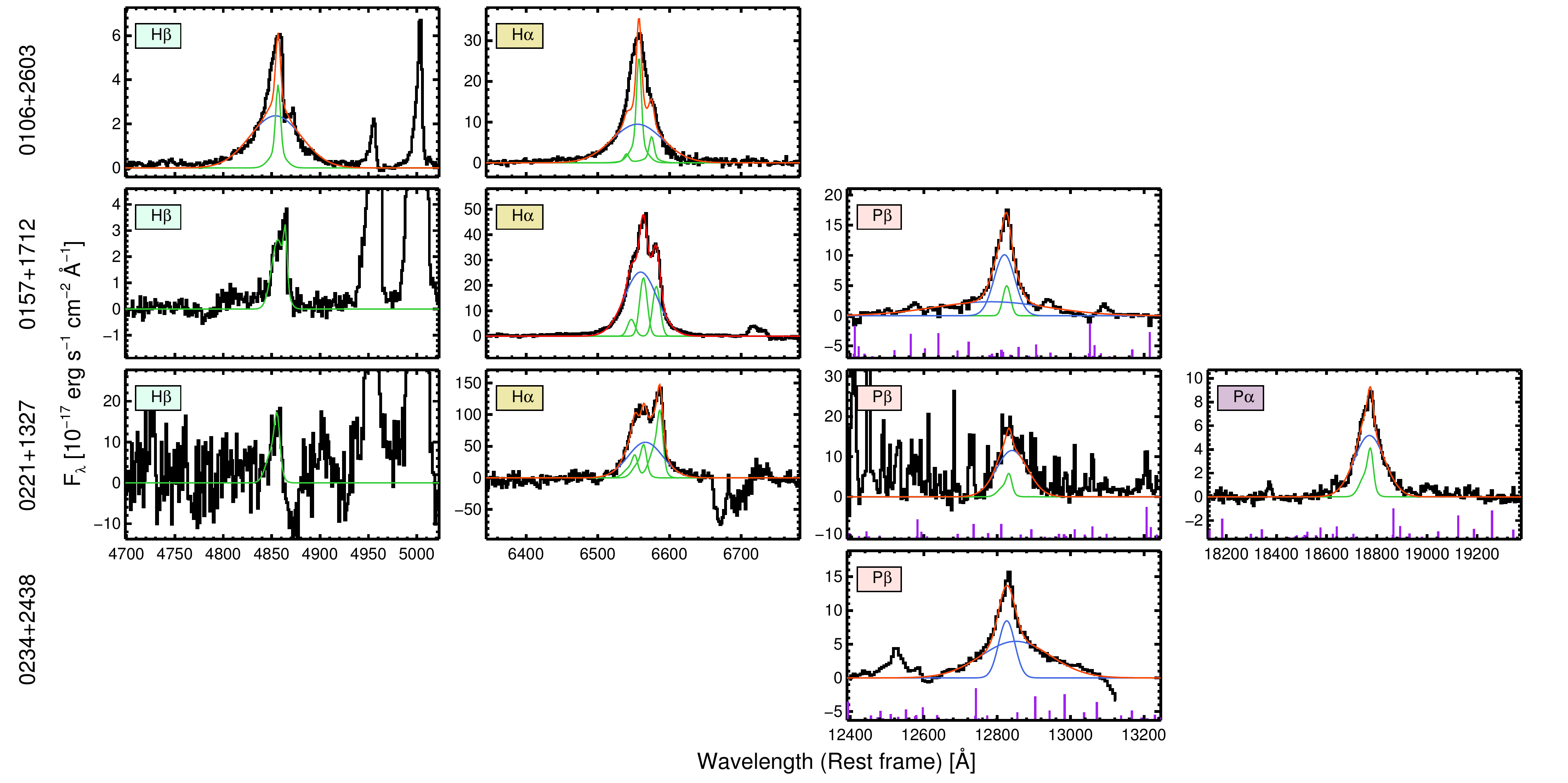}\\
	\includegraphics[width=\textwidth]{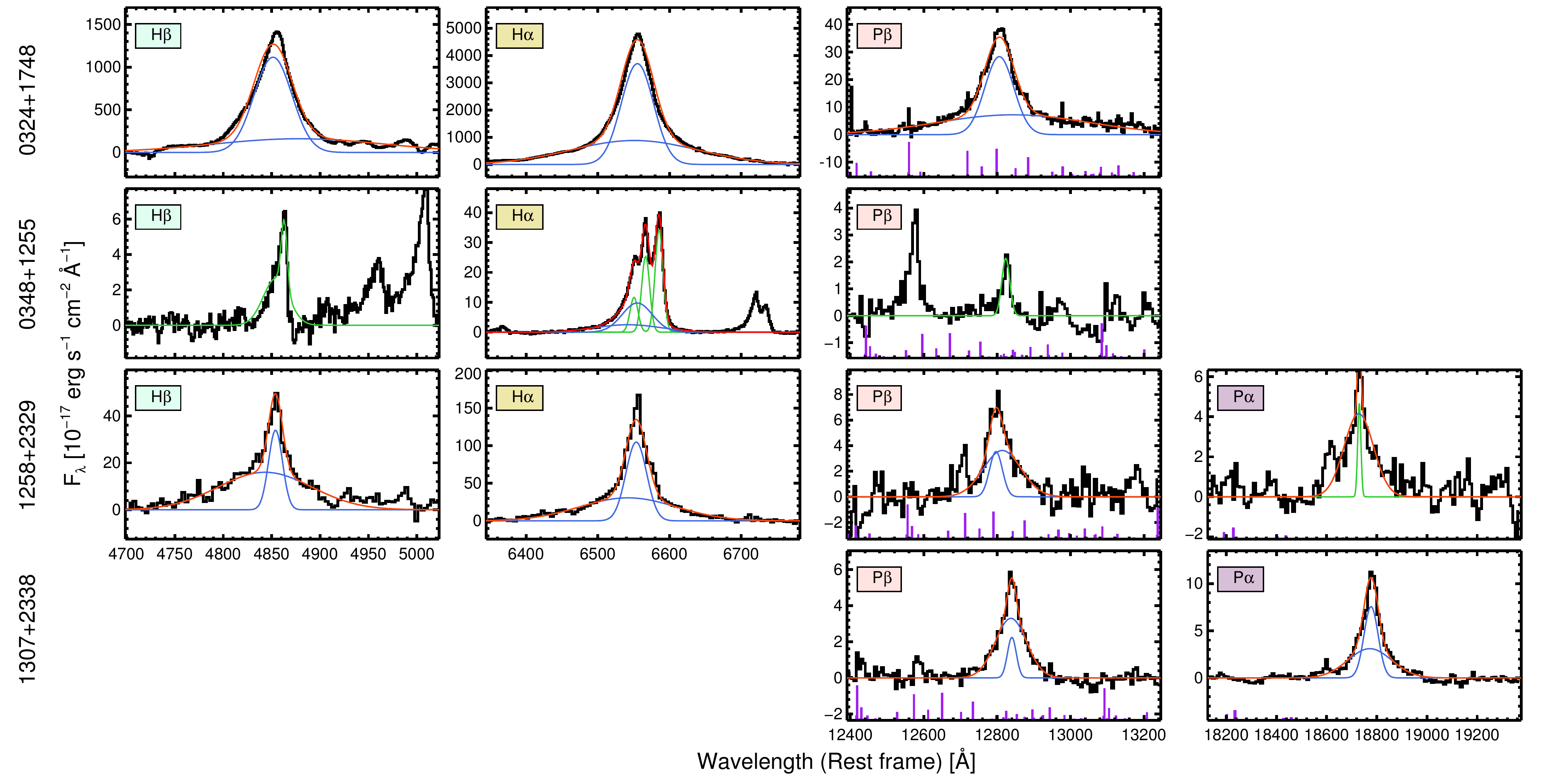}\\
	\caption{Results of the fitting of the H$\beta$, H$\alpha$, P$\beta$, and P$\alpha$ lines.
		The black lines indicate the continuum-subtracted observed spectra in the rest-frame.
		The red lines represent the best-fit model, and
		the green and blue lines represent the narrow and broad components, respectively.
		Moreover, for the P$\beta$ and P$\alpha$ figures, the purple lines at the bottom show the sky OH emission lines.}
\end{figure*}

\begin{figure*}
	\centering
	\figurenum{5}
	\includegraphics[width=\textwidth]{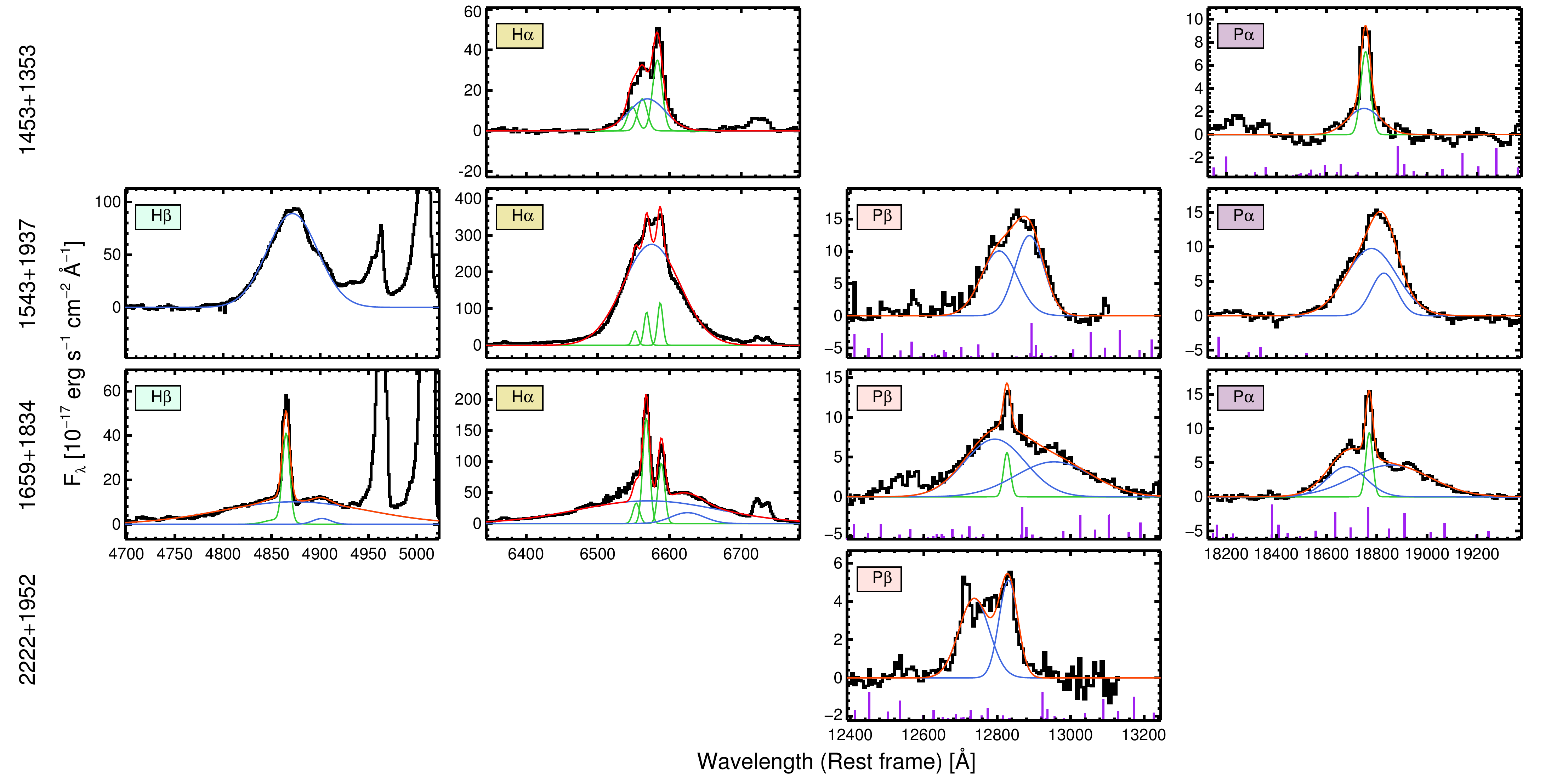}\\
	\includegraphics[width=\textwidth]{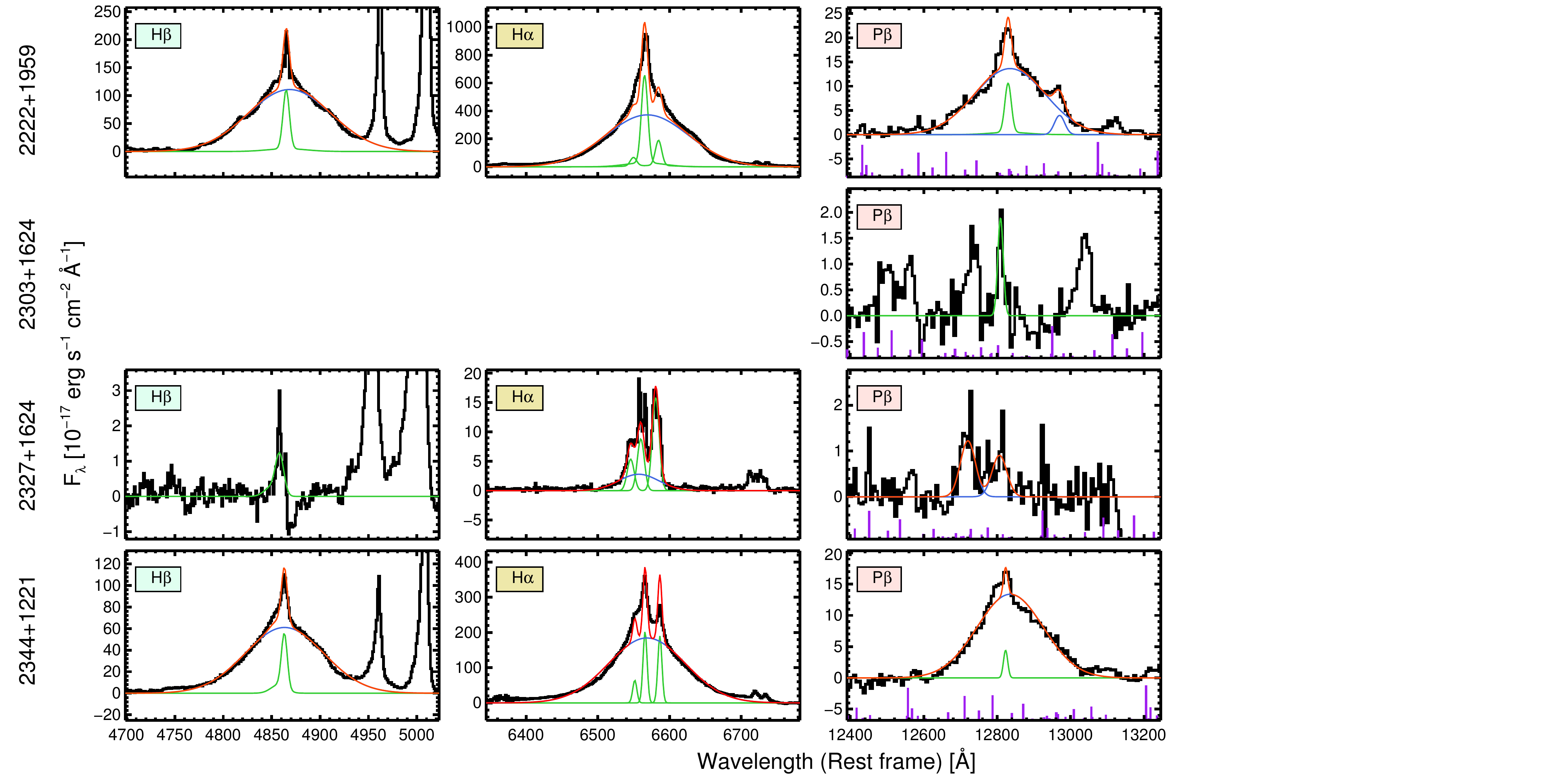}\\
	\caption{Continued}
\end{figure*}

 Strong correlations between the FWHMs of the Balmer and Paschen lines
 have been established for unobscured type 1 quasars \citep{landt08,kim10}.
 A tight correlation between the two quantities for our sample would imply that
 the luminosities originate from the same BLR, and the contribution of the narrow component is negligible.
 As shown in Figure 6, the measured FWHMs of Paschen lines are similar to those of Balmer lines,
 and the correlations between the two quantities are similar to those of unobscured type 1 quasars.

 In total, we obtain the broad line-luminosities and FWHMs of 7 H$\beta$, 12 H$\alpha$, 12 P$\beta$, and 6 P$\alpha$ lines.
 The measured luminosities and FWHMs of the hydrogen lines for broad and narrow components are summarized in Table 5 and Table 6, respectively.

\section{Reddening}

 In this section, we assume that the red colors of NIR-red AGNs originate from
 the dust extinction in their host galaxies, as shown in several previous studies \citep{glikman07,urrutia08,urrutia09,kim18}.
 Hence, measuring the color excess, $E(B-V)$, is important for investigating the intrinsic, i.e., un-reddened properties of dusty AGNs.
 In the following subsections,
 $E(B-V)$ values are derived using two methods,
 comparison of line-luminosity ratios and continuum slopes
 between unobscured and NIR-red AGNs.
 In this section, we use the reddening law, $k(\lambda)$, of \cite{fitzpatrick99},
 based on the Galactic extinction curve from 1000\,$\rm \AA{}$ to 3.5\,$\mu$m with $R_V=3.1$.

\subsection{Reddening derived from line-luminosity ratios}

 We measure the reddening from line-luminosity ratios ($E(B-V)_{\rm line}$) of NIR-red AGNs
 by using four broad line-luminosity ratios of Balmer to Paschen lines
 ($L_{\rm H\beta}$/$L_{\rm P\beta}$, $L_{\rm H\alpha}$/$L_{\rm P\beta}$,
 $L_{\rm H\beta}$/$L_{\rm P\alpha}$, and $L_{\rm H\alpha}$/$L_{\rm P\alpha}$).
 We use the correlation between the Balmer and Paschen line luminosities of 37 low-redshift ($ z < 0.5$) and bright ($J < 14$ or $K < 14.5$\,mag)
 unobscured type 1 quasars adopted from \cite{kim10}
 as their intrinsic line-luminosity ratios.
 By comparing the line-luminosity ratios, the $E(B-V)_{\rm line}$ values can be measured for 10 out of the 16 NIR-red AGNs.

\begin{figure*}
	\centering
	\figurenum{6}
	\includegraphics[scale=0.4]{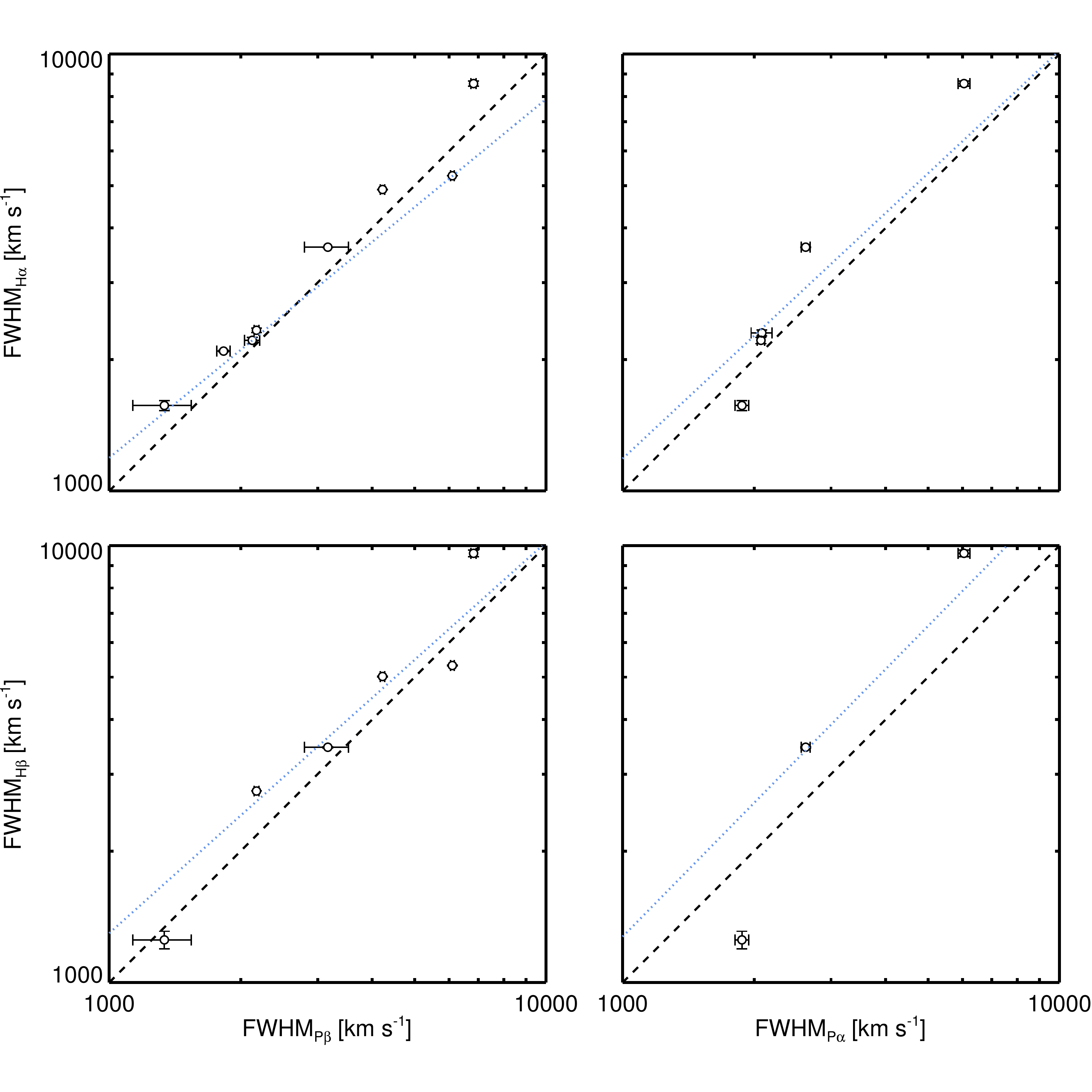}\\
	\caption{Comparisons of the FWHMs of the Balmer and Paschen lines.
		The black dashed lines show the two quantities are identical,
		and the blue dotted lines denote the adopted correlation of unobscured type 1 quasars from \cite{kim10}.}
\end{figure*}

 The $E(B-V)_{\rm line}$ values are computed by varying the amount of dust-reddening to minimize $\chi^2$,
 which is a function of the line-luminosity ratios of NIR-red AGNs
 ($R_{\rm obs,i,j} = L_{\rm obs, \lambda j} / L_{\rm obs, \lambda i}$)
 and unobscured type 1 quasars ($R_{\rm int,i,j} = L_{\rm int, \lambda j} / L_{\rm int, \lambda i}$), expressed as
\begin{equation}
 \chi^2 = \sum_{\rm i,j = 1}^{ N } \frac{(R_{\rm obs,i,j} - E (R_{\rm int,i,j}))^2}{\sigma_{\rm i,j}^2} .
\end{equation}
 Here, $N$ is the number of line-luminosity ratios,
 $\sigma_{\rm i,j}$ are the combined uncertainties of the line-luminosity ratios and the adopted correlations from \cite{kim10},
 and $E$ is a function for the dust-reddening expressed as
\begin{equation}
 \log \left( \frac{E(R_{\rm int,i,j})}{R_{\rm int,i,j}} \right) =\frac{E(B-V)}{1.086}(k({\rm \lambda i})-k({\rm \lambda j})).
\end{equation}

 For estimating the uncertainty of $E(B-V)_{\rm line}$, we perform 1000 Monte-Carlo simulations.
 We calculate new line-luminosity ratios by adding the measurement uncertainties of the line luminosities randomly
 to the observed line luminosities.
 The standard deviation of the 1000 newly measured $E(B-V)_{\rm line}$ values
 is taken as the uncertainty of the $E(B-V)_{\rm line}$.

 In Figure 7, we compare the observed and the dust-extinction-corrected line luminosities with the $E(B-V)_{\rm line}$ values.
 The measured $E(B-V)_{\rm line}$ values and uncertainties for the 10 NIR-red AGNs are summarized in Table 7.

\subsection{Reddening derived from continuum slopes}

 We measure the color excess values from continuum slope ($E(B-V)_{\rm cont}$) of the NIR-red AGNs by comparing
 the observed spectrum, $f(\lambda)$, to a model spectrum.
 The model spectrum combines a reddened quasar composite, $Q(\lambda)$, and a reddened stellar template, $S(\lambda)$.
 The intrinsic quasar composite, $Q_{0}(\lambda)$, is adopted from \cite{glikman06},
 which is a composition of an optical quasar composite \citep{brotherton01} and an NIR quasar composite \citep{glikman06}.
 They used unobscured type 1 quasars for constructing the optical and NIR quasar composites.
 For the intrinsic stellar template, $S_{0}(\lambda)$,
 we use K (MJD=51816, plate=396, and fiber=605), F (MJD=51990, plate=289, and fiber=5),
 and G (MJD=51957, plate=273, and fiber=304) type stellar spectra adopted from SDSS,
 since the K, F, and G type stars are the most dominant populations of the stellar composite template for NIR-red AGNs \citep{canalizo12}.

 In order to fit the $E(B-V)_{\rm cont}$ values,
 we fit the model spectrum to the observed spectrum, and the fitting function has a form of
\begin{equation}
 f(\lambda)=Q(\lambda) + S(\lambda).
\end{equation}
 Here, $Q(\lambda)$ and $S(\lambda)$ are the reddened spectra of $Q_{0}(\lambda)$ and $S_{0}(\lambda)$, respectively,
 with their $E(B-V)$ values as
\begin{equation}
 \log \left( \frac{X(\lambda)}{X_{0}(\lambda)} \right) =-\frac{k(\lambda) E(B-V)_{X}}{1.086},
\end{equation}
 where $E(B-V)_{Q}$ is taken as $E(B-V)_{\rm cont}$.
 Here, $X(\lambda)$ denotes $Q(\lambda)$ or $S(\lambda)$, and
 $X_{0}(\lambda)$ is $Q_{0}(\lambda)$ or $S_{0}(\lambda)$.

 For the fit, we use only a limited wavelength range (3790--10000\,$\rm \AA{}$),
 because \cite{glikman07} reported that fitting with the optical and NIR combined spectrum yields
 extremely poor results for one-third of red AGNs.
 Moreover, to exclude strong emission lines,
 wavelength regions of 3790--4700, 5100-6400, and 6700--10000\,$\rm \AA{}$ are used.
 There are remaining moderate emission lines
 (e.g., \ion{He}{1} $\lambda$3889, H$\epsilon$ $\lambda$3970, H$\delta$ $\lambda$4102, H$\gamma$ $\lambda$4340,
 \ion{He}{1} $\lambda$5876, [\ion{O}{1}] $\lambda\lambda$6300, 6364, and \ion{O}{1} $\lambda$8447),
 but the effects on the fit are negligible.
 In order to find the most likely stellar template,
 we calculate $\chi^2$ values for the fits with the K, F, and G type star spectra,
 and the fit with the minimum $\chi^2$ is used as the best fit.
 From this fitting procedure, we measure the $E(B-V)_{\rm cont}$ values for 12 NIR-red AGNs,
 and Figure 8 shows the fitting results.
 The measured $E(B-V)_{\rm cont}$ values and uncertainties are summarized in Table 7.

\subsection{Discussion for the two types of reddening}

 We compare the $E(B-V)_{\rm line}$ values to the $E(B-V)_{\rm cont}$ values for the 10 NIR-red AGNs
 that have both $E(B-V)_{\rm line}$ and $E(B-V)_{\rm cont}$ in Figure 9.
 The two types of $E(B-V)$ values are consistent,
 but there is a weak trend of $E(B-V)_{\rm cont} - E(B-V)_{\rm line} \sim {0.281}$.
 We estimate the Pearson correlation coefficient between the two quantities.
 For estimating the coefficient, we exclude 0324$+$1748, which has negative values for both types of $E(B-V)$,
 and assume that the negative $E(B-V)$ values ($E(B-V)_{\rm line}$ values of 1258$+$2329 and 2222$+$1959) are 0.
 The measured coefficient is 0.911, and the rms scatter with respect to a one-to-one correlation is 0.223.
 This result supports that the two measurements of $E(B-V)$ are mutually verified.

\begin{deluxetable*}{ccccccccccccc}
	\tablecolumns{13}
	\tablewidth{0pt}
	\tablenum{5}
	\tablecaption{Hydrogen line measurements for the broad component \label{tbl5}}
	\tablehead{
		\colhead{Object Name}&	\colhead{}&
		\multicolumn{2}{c}{H$\beta$}&	\colhead{}&
		\multicolumn{2}{c}{H$\alpha$}&	\colhead{}&
		\multicolumn{2}{c}{P$\beta$}&	\colhead{}&
		\multicolumn{2}{c}{P$\alpha$}\\
		\cline{3-4} \cline{6-7} \cline{9-10} \cline{12-13}\\
		\colhead{}&	\colhead{}&	
		\colhead{FWHM}&	\colhead{$L$}&	\colhead{}&
		\colhead{FWHM}&	\colhead{$L$}&	\colhead{}&
		\colhead{FWHM}&	\colhead{$L$}&	\colhead{}&
		\colhead{FWHM}&	\colhead{$L$}\\
		\colhead{}&	\colhead{}&
		\colhead{($\rm{km\,s^{-1}}$)}&	\colhead{($\rm{10^{40}\,erg\,s^{-1}}$)}&	\colhead{}&
		\colhead{($\rm{km\,s^{-1}}$)}&	\colhead{($\rm{10^{40}\,erg\,s^{-1}}$)}&	\colhead{}&
		\colhead{($\rm{km\,s^{-1}}$)}&	\colhead{($\rm{10^{40}\,erg\,s^{-1}}$)}&	\colhead{}&
		\colhead{($\rm{km\,s^{-1}}$)}&	\colhead{($\rm{10^{40}\,erg\,s^{-1}}$)}
	}
	
	\startdata
	0106$+$2603&	&	3281$\pm$5& 	1.000$\pm$0.003&	&	3227$\pm$8&		5.312$\pm$0.028&	&	--&				--&					&	
	--&				--\\
	0157$+$1712&	&	--&				--&					&	2090$\pm$7&		2.009$\pm$0.017&	&	1827$\pm$63&	2.133$\pm$0.102&	&	
	--&				--\\
	0221$+$1327&	&	--&				--&					&	2212$\pm$36&	1.875$\pm$0.070&	&	2125$\pm$84&	0.682$\pm$0.043&	&	
	2073$\pm$40&	0.436$\pm$0.013\\
	0234$+$2438&	&	--&				--&					&	--&				--&					&	1515$\pm$77&	5.735$\pm$0.202&	&
	--&				--\\
	0324$+$1748&	&	2744$\pm$0&		339.5$\pm$1.6&		&	2333$\pm$28&	1513$\pm$3&			&	2173$\pm$28&	24.67$\pm$0.42&		&
	--&				--\\
	0348$+$1255&	&	--&				--&					&	2539$\pm$325&	1.027$\pm$0.038&	&	--&				--&					&
	--&				--\\
	1258$+$2329&	&	1252$\pm$58&	5.601$\pm$0.137&	&	1569$\pm$41&	19.75$\pm$0.39&		&	1337$\pm$204&	1.368$\pm$0.152&	&
	1876$\pm$67&	1.342$\pm$0.084\\
	1307$+$2338&	&	--&				--&					&	--&				--&					&	1125$\pm$106&	1.040$\pm$0.073&	&
	1113$\pm$35&	2.997$\pm$0.129\\
	1453$+$1353&	&	--&				--&					&	2301$\pm$35&	0.541$\pm$0.033&	&	--&				--&					&
	2083$\pm$114&	0.201$\pm$0.017\\
	1543$+$1937&	&	3456$\pm$10&	9.554$\pm$0.038&	&	3615$\pm$4&		44.27$\pm$0.10&		&	3165$\pm$365&	3.976$\pm$0.379&	&
	2624$\pm$62&	5.254$\pm$0.193\\
	1659$+$1834&	&	9603$\pm$171&	1.603$\pm$0.023&	&	8561$\pm$120&	8.117$\pm$0.051&	&	6822$\pm$90&	2.128$\pm$0.036&	&
	6045$\pm$187&	2.135$\pm$0.058\\
	2222$+$1952&	&	--&				--&					&	--&				--&					&	3212$\pm$502&	3.647$\pm$0.201&	&
	--&				--\\
	2222$+$1959&	&	5316$\pm$9&		16.28$\pm$0.04&		&	5268$\pm$4&		73.07$\pm$0.12&		&	6099$\pm$60&	3.705$\pm$0.069&	&
	--&				--\\
	2303$+$1624&	&	--&				--&					&	--&				--&					&	--&				--&					&
	--&				--\\
	2327$+$1624&	&	--&				--&					&	2554$\pm$71&	0.940$\pm$0.050&	&	--&				0.450$\pm$0.093&	&
	--&				--\\
	2344$+$1221&	&	5018$\pm$13&	7.423$\pm$0.030&	&	4899$\pm$5&		29.56$\pm$0.07&		&	4223$\pm$50&	3.273$\pm$0.054&	&
	--&				--
	\enddata
	\tablecomments{The listed fluxes of H$\beta$, H$\alpha$, P$\beta$, and P$\alpha$ lines are not corrected
		from the dust extinction caused by their host galaxies.}
\end{deluxetable*}

\begin{deluxetable*}{ccccccccccccc}
	\tablecolumns{13}
	\tablewidth{0pt}
	\tablenum{6}
	\tablecaption{Hydrogen line measurements for the narrow component \label{tbl6}}
	\tablehead{
		\colhead{Object Name}&	\colhead{}&
		\multicolumn{2}{c}{H$\beta$}&	\colhead{}&
		\multicolumn{2}{c}{H$\alpha$}&	\colhead{}&
		\multicolumn{2}{c}{P$\beta$}&	\colhead{}&
		\multicolumn{2}{c}{P$\alpha$}\\
		\cline{3-4} \cline{6-7} \cline{9-10} \cline{12-13}\\
		\colhead{}&	\colhead{}&	
		\colhead{$L$}&	\colhead{FWHM}&	\colhead{}&
		\colhead{$L$}&	\colhead{FWHM}&	\colhead{}&
		\colhead{$L$}&	\colhead{FWHM}&	\colhead{}&
		\colhead{$L$}&	\colhead{FWHM}\\
		\colhead{}&	\colhead{}&
		\colhead{($\rm{10^{38}\,erg\,s^{-1}}$)}&	\colhead{($\rm km\,s^{-1}$)}&	\colhead{}&
		\colhead{($\rm{10^{38}\,erg\,s^{-1}}$)}&	\colhead{($\rm km\,s^{-1}$)}&	\colhead{}&
		\colhead{($\rm{10^{38}\,erg\,s^{-1}}$)}&	\colhead{($\rm km\,s^{-1}$)}&	\colhead{}&
		\colhead{($\rm{10^{38}\,erg\,s^{-1}}$)}&	\colhead{($\rm km\,s^{-1}$)}
	}
	
	\startdata
	0106$+$2603&	&	20.17$\pm$0.03& 	359.9$\pm$36.2&	&	140.6$\pm$0.8&		359.9$\pm$36.2&	&	--&					--&				&	
	--&					--\\
	0157$+$1712&	&	6.869$\pm$0.300&	988.4$\pm$63.8&	&	40.24$\pm$0.71&		563.7$\pm$20.7&	&	17.18$\pm$2.95&		563.7$\pm$20.7&	&	
	--&					--\\
	0221$+$1327&	&	10.73$\pm$4.40&		539.5$\pm$54.9&	&	32.92$\pm$1.62&		539.5$\pm$54.9&	&	3.615$\pm$0.656&	539.5$\pm$54.9&	&	
	2.566$\pm$0.124&	539.5$\pm$54.9\\
	0234$+$2438&	&	--&					--&				&	--&					--&				&	--&					--&				&
	--&					--\\
	0324$+$1748&	&	--&					--&				&	--&					--&				&	--&					--&				&
	--&					--\\
	0348$+$1255&	&	14.88$\pm$0.10&		718.7$\pm$25.7&	&	45.51$\pm$0.60&		595.8$\pm$6.8&	&	7.530$\pm$0.776&	595.8$\pm$6.8&	&
	--&					--\\
	1258$+$2329&	&	--&					--&				&	--&					--&				&	--&					--&				&
	15.77$\pm$4.82&		190.3$\pm$36.1\\
	1307$+$2338&	&	--&					--&				&	--&					--&				&	--&					--&				&
	--&					--\\
	1453$+$1353&	&	--&					--&				&	13.76$\pm$0.87&		695.5$\pm$47.5&	&	--&					--&				&
	17.55$\pm$0.65&		695.5$\pm$47.5\\
	1543$+$1937&	&	--&					--&				&	128.0$\pm$2.1&		387.2$\pm$10.8&	&	--&					--&				&
	--&					--\\
	1659$+$1834&	&	35.11$\pm$9.53&		628.7$\pm$22.1&	&	161.5$\pm$2.0&		501.2$\pm$5.1&	&	10.57$\pm$0.65&		501.2$\pm$5.1&	&
	26.17$\pm$1.27&		501.2$\pm$5.1\\
	2222$+$1952&	&	--&					--&				&	--&					--&				&	--&					--&				&
	--&					--\\
	2222$+$1959&	&	144.6$\pm$1.3&		538.9$\pm$41.8&	&	875.9$\pm$5.3&		538.9$\pm$41.8&	&	11.71$\pm$0.86&		538.9$\pm$41.8&	&
	--&					--\\
	2303$+$1624&	&	--&					--&				&	--&					--&				&	7.887$\pm$3.145&	397.4$\pm$115.9&	&
	--&					--\\
	2327$+$1624&	&	7.587$\pm$9.783&	719.6$\pm$77.7&	&	49.27$\pm$1.45&		518.9$\pm$33.1&	&	--&					--&				&
	--&					--\\
	2344$+$1221&	&	60.26$\pm$0.91&		449.3$\pm$30.9&	&	166.5$\pm$2.7&		301.1$\pm$10.0&	&	6.810$\pm$1.109&	301.1$\pm$10.0&	&
	--&					--
	\enddata
	\tablecomments{The listed fluxes are not corrected from the dust extinction caused by their host galaxies.}
\end{deluxetable*}

 Moreover, we compare the two types of $E(B-V)$ values to the $E(B-V)$ values adopted from Canalizo et al. (2012; hereafter, $E(B-V)_{\rm C12}$),
 and this comparison is shown in Figure 9.
 They measured the $E(B-V)_{\rm C12}$ by comparing the observed continuum spectrum to
 the SDSS composite QSO spectrum \citep{berk01} reddened with a Small Magellanic Cloud reddening law \citep{bouchet85},
 and five NIR-red AGNs (0157$+$1712, 0221$+$1327, 0348$+$1255, 1659$+$1834, and 2327$+$1624) are overlapped with our sample.
 They showed that the $E(B-V)_{\rm C12}$ values were generally consistent with the $E(B-V)$ values derived by using Balmer decrements,
 and the difference between this two quantities was $\sim$0.3.
 Our $E(B-V)_{\rm cont}$ and $E(B-V)_{\rm line}$ values are generally but somewhat weakly consistent with the $E(B-V)_{\rm C12}$ values.
 Between the $E(B-V)_{\rm cont}$ and $E(B-V)_{\rm C12}$ values, the Pearson correlation coefficient is 0.579, and the rms scatter is 0.484.
 For the $E(B-V)_{\rm line}$ values, the result is generally same as
 the coefficient is 0.663 with the rms scatter of 0.582.
 We found a trend that the $E(B-V)$ values from this work is less than the $E(B-V)_{\rm C12}$ values,
 as much as $\Delta E(B-V) \sim 0.520$, but this trend is not significant due to small number statistics.

 Unlike our results, previous studies \citep{glikman07,kim18} reported that
 the two types of $E(B-V)$ values are far from a one-to-one correlation.
 The Pearson correlation coefficient between the two quantities is only -0.21, with an rms scatter of 0.68 \citep{kim18}.

 In the previous studies, the $E(B-V)_{\rm cont}$ values are from \cite{glikman07} and \cite{urrutia09},
 and the $E(B-V)_{\rm line}$ values are from \cite{glikman07}.
 To obtain the $E(B-V)_{\rm cont}$ values,
 they fit the continua using the quasar component only, without the stellar component.
 In this study, considering the continuum spectra of the most of NIR-red AGNs are dominated by the quasar component,
 the measurement technique for the $E(B-V)_{\rm cont}$ is almost the same, and
 it is hard to believe that the contrasting result comes from the discrepancy of the $E(B-V)_{\rm cont}$ values.

\begin{figure*}
	\centering
	\figurenum{7}
	\includegraphics[scale=0.5]{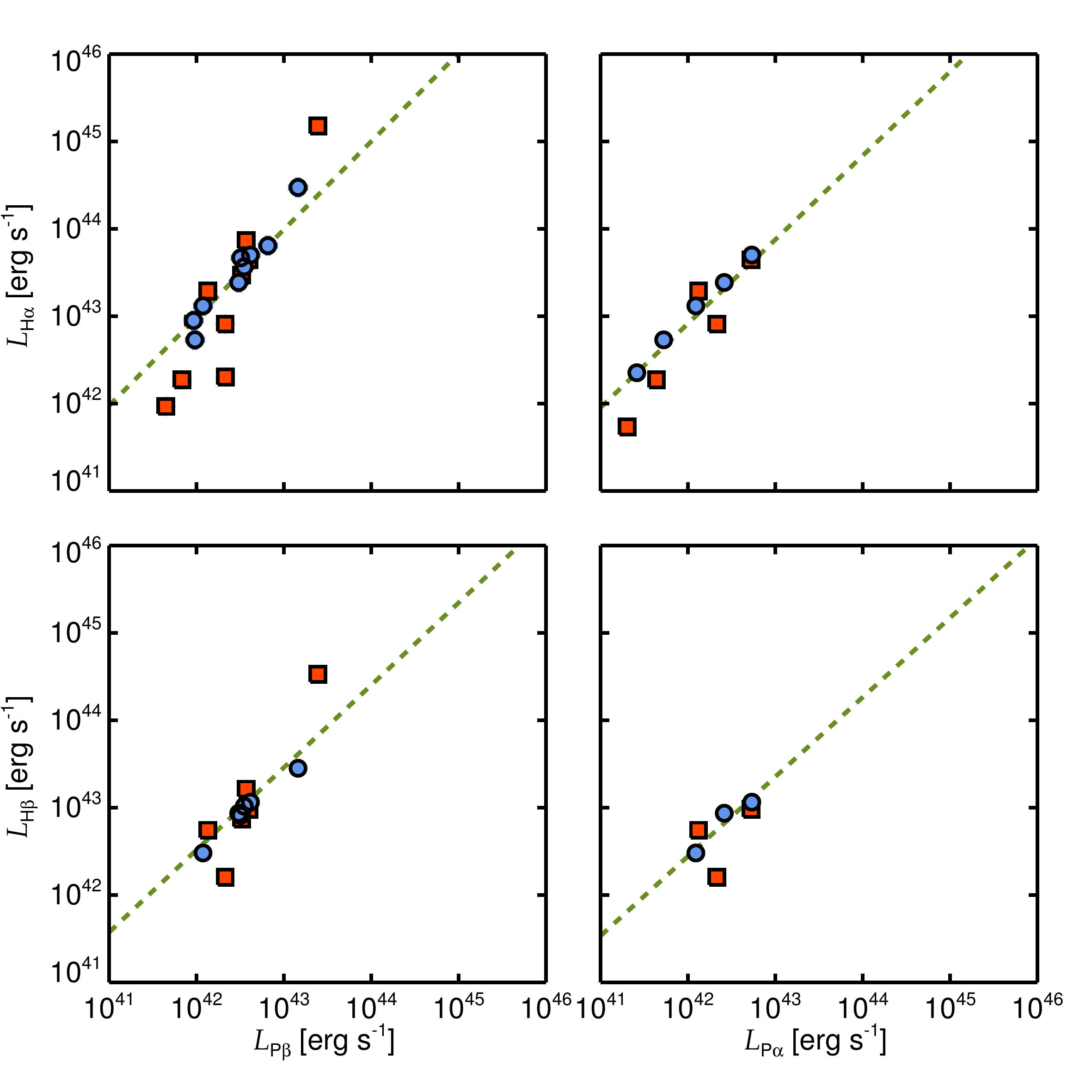}\\
	\caption{Observed and dust-extinction-corrected line luminosities of NIR-red AGNs.
		The red squares show the observed line luminosities,
		and the blue circles mean the dust-extinction-corrected line luminosities with the measured $E(B-V)_{\rm line}$ values.
		The green dotted lines represent the correlations between the Balmer and Paschen line luminosities \citep{kim10},
		and these correlations are used for estimating the $E(B-V)_{\rm line}$ values.}
\end{figure*}

 However, in order to measure the $E(B-V)_{\rm line}$ values,
 Glikman et al. used somewhat different way.
 They measured the $E(B-V)_{\rm line}$ values using Balmer decrements (hereafter, $E(B-V)_{\rm BD1}$).
 The $E(B-V)_{\rm BD1}$ values were obtained by following formula:
\begin{equation}
 E(B-V)_{\rm BD1}= \frac{1.086}{k({\rm H\beta})-k({\rm H\alpha})} \ln \left( \frac{[F_{\rm H\alpha}/F_{\rm H\beta}]_{\rm measured}}{[F_{\rm H\alpha}/F_{\rm H\beta}]_{\rm FBQS}} \right).
\end{equation}
 Here, $k({\rm X})$ is the extinction law of \cite{calzetti94} at the wavelength of the X line,
 and $[F_{\rm H\alpha}/F_{\rm H\beta}]_{\rm measured}$ and $[F_{\rm H\alpha}/F_{\rm H\beta}]_{\rm FBQS}$
 are the $F_{\rm H\alpha}/F_{\rm H\beta}$ from the spectra of red quasars and 
 the Faint Images of the Radio Sky at Twenty-Centimeters (FIRST) Bright Quasar Survey (FBQS; \citealt{gregg96}) composite spectrum, respectively.
 In this equation, the $[F_{\rm H\alpha}/F_{\rm H\beta}]_{\rm FBQS}$ is used as the
 intrinsic $F_{\rm H\alpha}/F_{\rm H\beta}$ of red quasars,
 which is a fixed value of 4.526 when only the broad component is treated \citep{glikman07}.

 Using this technique, we can measure the $E(B-V)_{\rm BD1}$ values for seven NIR-red AGNs,
 and they are summarized in Table 7.
 We compare the $E(B-V)_{\rm BD1}$ values to the $E(B-V)_{\rm line}$ and $E(B-V)_{\rm cont}$ values in Figure 10.
 For the $E(B-V)_{\rm BD1}$ values versus $E(B-V)_{\rm line}$ values, six NIR-red AGNs are used,
 and the measured Pearson correlation coefficient is 0.314 with an rms scatter of 0.257.
 For the comparison between the $E(B-V)_{\rm BD1}$ and $E(B-V)_{\rm cont}$ values,
 we used seven NIR-red AGNs that have both quantities,
 and the Pearson coefficient and the rms scatter are 0.709 and 0.255, respectively.

 In order to figure out what makes this difference,
 we measure different types of Balmer-decrement-based $E(B-V)$ values (hereafter, $E(B-V)_{\rm BD2}$).
 First, we combine two relations of $L_{\rm H\alpha}$--$L_{\rm P\alpha}$ and $L_{\rm H\beta}$--$L_{\rm P\alpha}$
 from \cite{kim10} to make a relation of $L_{\rm H\alpha}$--$L_{\rm H\beta}$.
 The combined relation is
\begin{equation}
\log \left( \frac{L_{\rm H\alpha}}{{\rm 10^{42}\,erg\,s^{-1}}} \right) =
0.509 + 1.056 \log \left( \frac{L_{\rm H\beta}}{{\rm 10^{42}\,erg\,s^{-1}}} \right)
\end{equation}
 that is used as the intrinsic $L_{\rm H\alpha}$--$L_{\rm H\beta}$ relation of NIR-red AGNs.
 Second, we measure the $E(B-V)_{\rm BD2}$ values by varying the amount of dust-reddening to minimize $\chi^2$,
 which is a function of the $L_{\rm H\alpha}$/$L_{\rm H\beta}$ of NIR-red AGNs ($R_{\rm obs}$)
 and unobscured type 1 quasars ($R_{\rm int}$), expressed as
\begin{equation}
\chi^2 = \frac{(R_{\rm obs}-E(R_{\rm int}))^2}{\sigma^2}.
\end{equation}
 Here, the $R_{\rm int}$ is derived from the above relation of $L_{\rm H\alpha}$--$L_{\rm H\beta}$,
 $E$ is a dust-reddening function, and $\sigma$ is the combined uncertainty of the $R_{\rm obs}$ and $R_{\rm int}$.
 The measured $E(B-V)_{\rm BD2}$ values are summarized in Table 7.

 The $E(B-V)_{\rm BD2}$ values show tighter correlations with
 the $E(B-V)_{\rm line}$ and $E(B-V)_{\rm cont}$ values than the $E(B-V)_{\rm BD1}$ values,
 but these comparisons cannot be meaningful due to the small number statistics.
 The comparisons of the $E(B-V)_{\rm BD2}$ values with
 the $E(B-V)_{\rm line}$ and $E(B-V)_{\rm cont}$ values are shown in Figure 10.
 Between the $E(B-V)_{\rm BD2}$ and $E(B-V)_{\rm line}$ values,
 the Pearson correlation coefficient is 0.816, with an rms scatter of 0.249.
 For the $E(B-V)_{\rm BD2}$ and $E(B-V)_{\rm cont}$ values,
 the Pearson correlation coefficient and the rms scatter are 0.941 and 0.136, respectively.
 These Pearson correlation coefficients are significantly bigger than the coefficients from the $E(B-V)_{\rm BD1}$ values.

 Although the difference cannot be meaningful due to the small number statistics,
 if there is a difference, we suspect the different intrinsic $L_{\rm H\alpha}$/$L_{\rm H\beta}$ causes these conflicting results.
 Because the intrinsic Balmer decrements are fixed to 4.526 for deriving the $E(B-V)_{\rm BD1}$ values,
 these quantities vary with the $L_{\rm H\beta}$ values for the $E(B-V)_{\rm BD2}$ values.
 For example, when the $L_{\rm H\beta}$ is increased from $\rm 10^{42}\,erg\,s^{-1}$ to $\rm 10^{44}\,erg\,s^{-1}$,
 the intrinsic Balmer decrement increases from 3.23 to 4.18,
 which makes up $\sim$22\% of the discrepancy of the measured $E(B-V)$ values.

\begin{deluxetable*}{ccccc}
	\tabletypesize{\scriptsize}
	\tablecolumns{10}
	\tablewidth{0pt}
	\tablenum{7}
	\tablecaption{Four kinds of $E(B-V)$ values \label{tbl7}}
	\tablewidth{0pt}
	\tablehead{
		\colhead{Object}&	\colhead{$E(B-V)_{\rm line}$}&	\colhead{$E(B-V)_{\rm cont}$}&	\colhead{$E(B-V)_{\rm BD1}$}&	\colhead{$E(B-V)_{\rm BD2}$}\\
		\colhead{}&			\colhead{(mag)}&				\colhead{(mag)}&				\colhead{(mag)}&				\colhead{(mag)}}
	\startdata
	0106$+$2603&	--&					0.675$\pm$0.001&	0.137$\pm$0.005&	0.371$\pm$0.005\\
	0157$+$1712&	1.596$\pm$0.034&	2.157$\pm$0.008&	--&					--\\
	0221$+$1327&	0.484$\pm$0.032&	0.990$\pm$0.008&	--&					--\\
	0234$+$2438&	--&					--&					--&					--\\
	0324$+$1748&	-0.745$\pm$0.007&	-0.026$\pm$0.000&	-0.013$\pm$0.004&	-0.003$\pm$0.004\\
	0348$+$1255&	--&					1.326$\pm$0.010&	--&					--\\
	1258$+$2329&	-0.179$\pm$0.021&	0.091$\pm$0.001&	-0.213$\pm$0.027&	-0.006$\pm$0.023\\
	1307$+$2338&	--&					--&					--&					--\\
	1453$+$1353&	0.657$\pm$0.060&	0.890$\pm$0.002&	--&					--\\
	1543$+$1937&	0.057$\pm$0.013&	0.169$\pm$0.000&	0.020$\pm$0.004&	0.175$\pm$0.004\\
	1659$+$1834&	0.505$\pm$0.007&	0.636$\pm$0.001&	0.096$\pm$0.014&	0.315$\pm$0.012\\
	2222$+$1952&	--&					--&					--&					--\\
	2222$+$1959&	-0.208$\pm$0.008&	0.271$\pm$0.000&	-0.007$\pm$0.003&	0.129$\pm$0.002\\
	2303$+$1624&	--&					--&					--&					--\\
	2327$+$1624&	1.043$\pm$0.154&	0.681$\pm$0.003&	--&					--\\
	2344$+$1221&	0.101$\pm$0.008&	0.259$\pm$0.000&	-0.109$\pm$0.004&	0.073$\pm$0.004
	\enddata
\end{deluxetable*}

 \subsection{Color selection for dusty red AGNs}
 In \cite{urrutia09}, their red AGNs were classified to have $E(B-V) > 0.1$.
 Only two objects among the $\sim$50 candidates in \cite{urrutia09} have $E(B-V) < 0.1$,
 and these two were not classified as dusty red AGNs.
 In this study, considering the rms scatter of the $E(B-V)$ values,
 $E(B-V) \sim 0.2$ is virtually identical to no extinction,
 and we classify the objects with $E(B-V) > 0.2$ as dusty red AGNs.
 According to this criteria, among our sample, three (0324$+$1748, 1258$+$2329, and 1543$+$1937) or
 five (0324$+$1748, 1258$+$2329, 1543$+$1937, 2222$+$1959, and 2344$+$1221) objects
 cannot be classified as dusty red AGNs based on the $E(B-V)_{\rm cont}$ or $E(B-V)_{\rm line}$ values, respectively,
 and the fraction of the low-$E(B-V)$ red AGNs (LERA) is bigger than that of \cite{urrutia09}.

 To figure out why this discrepancy arises,
 we check the differences in sample selection for the two studies.
 In order to select the red AGN candidates,
 \cite{urrutia09} used the optical-NIR ($r'-K>5$) and NIR colors ($J-K>1.3$) of FIRST-detected objects,
 but our sample was selected by NIR color only ($J-K>2$).
 Our entire sample within the FIRST coverage has detections in this survey,
 so the difference in the LERA fraction of the two samples originates from the lack of the optical-NIR color selection,
 and the NIR color alone is not sufficient to select dusty red AGNs.
 As shown in Figure 11, the change in $E(B-V)$ due to the increase in $J-K$ is not significant,
 but the $E(B-V)$ values are larger ($E(B-V) > 0$) when $g'-K > 5$.
 Moreover, this result is supported by \cite{maddox06}, in which
 a significant fraction of unobscured type 1 AGNs at low redshifts have such an NIR color of $J-K > 2$.
 We conclude that a significant portion of the NIR-red AGNs is unobscured or only mildly obscured type 1 AGNs.

\begin{figure*}[!t]
	\centering
	\figurenum{8}
	\includegraphics[scale=0.35]{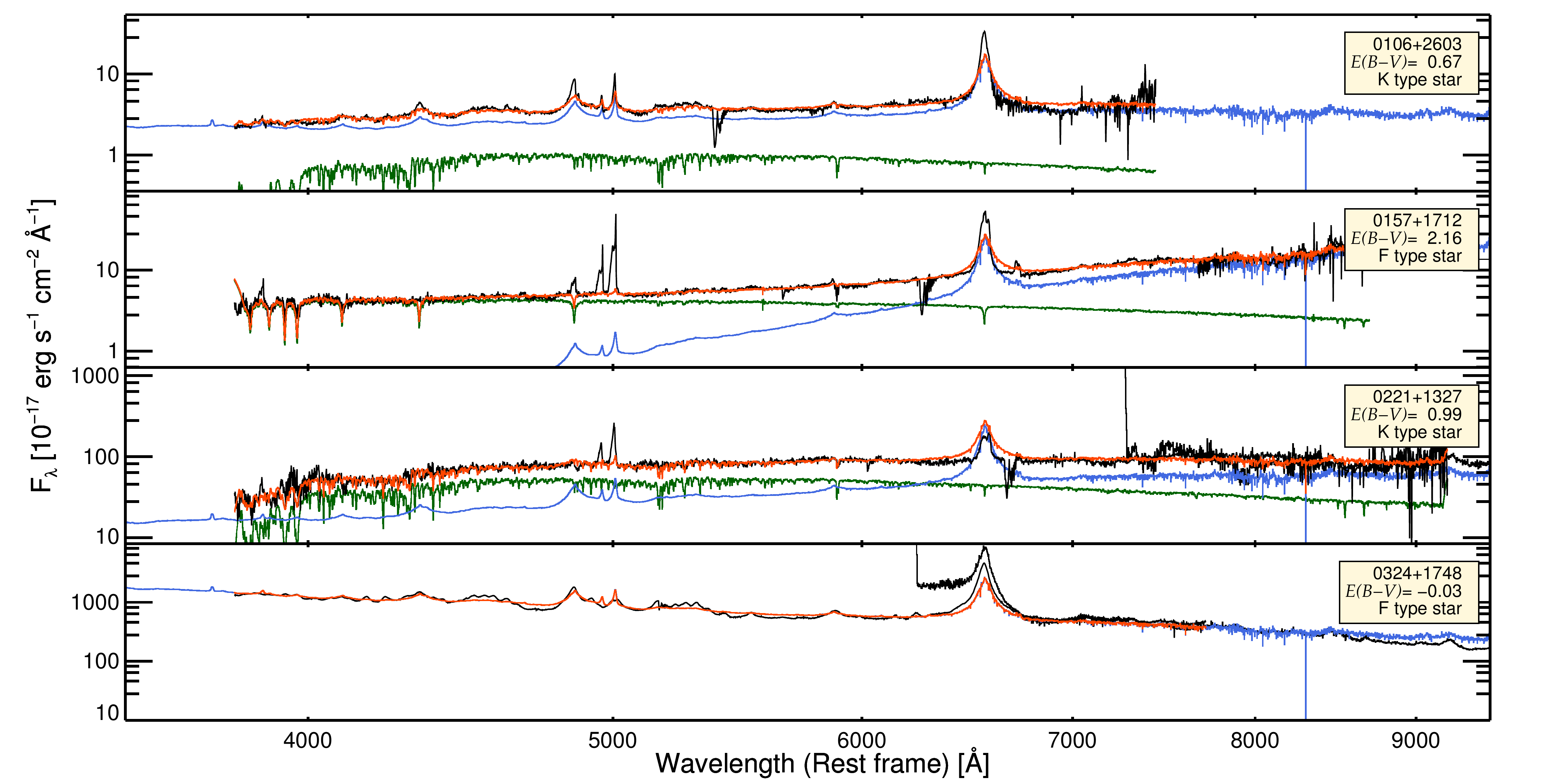}\\
	\includegraphics[scale=0.35]{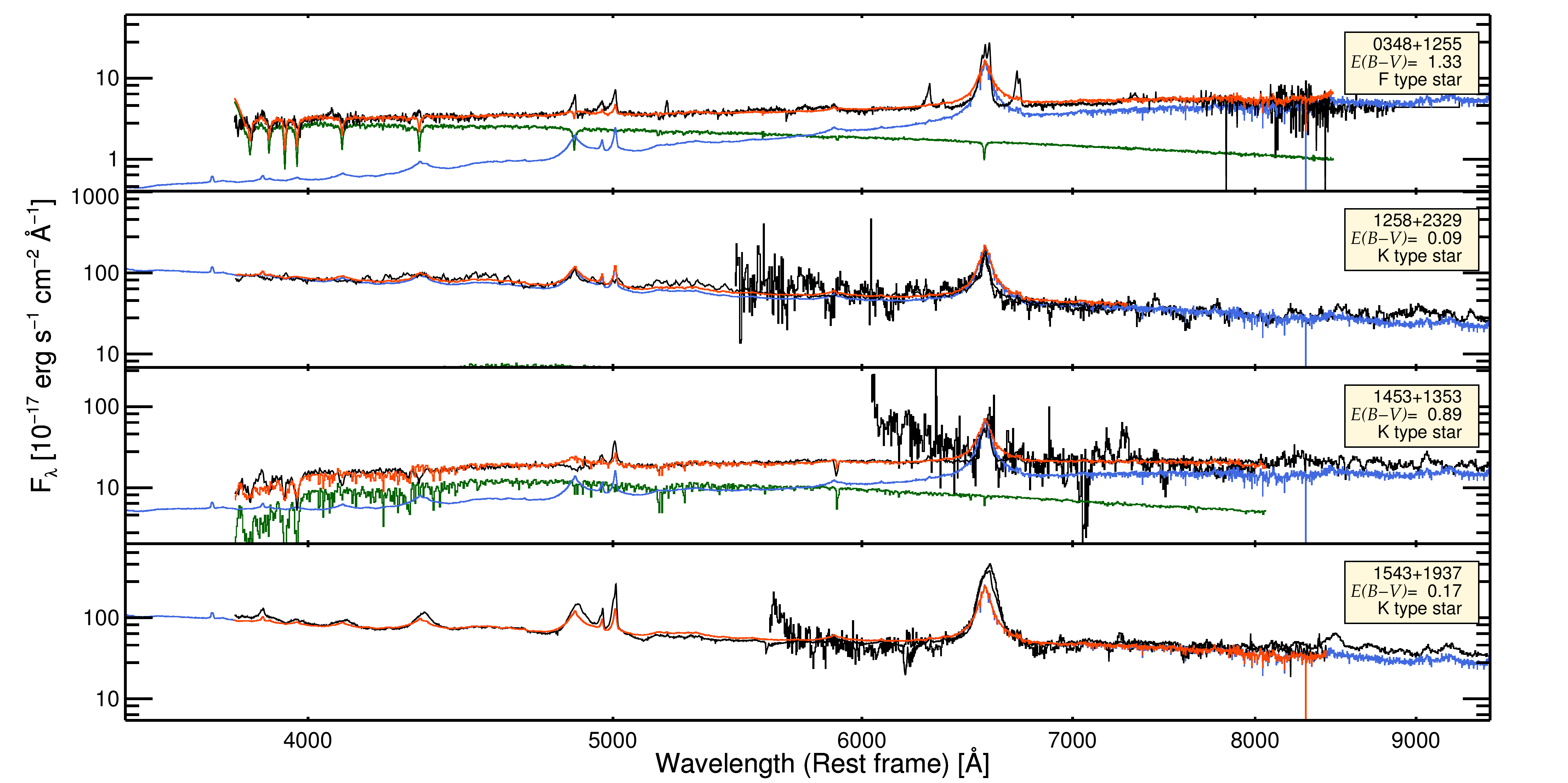}\\
	\includegraphics[scale=0.35]{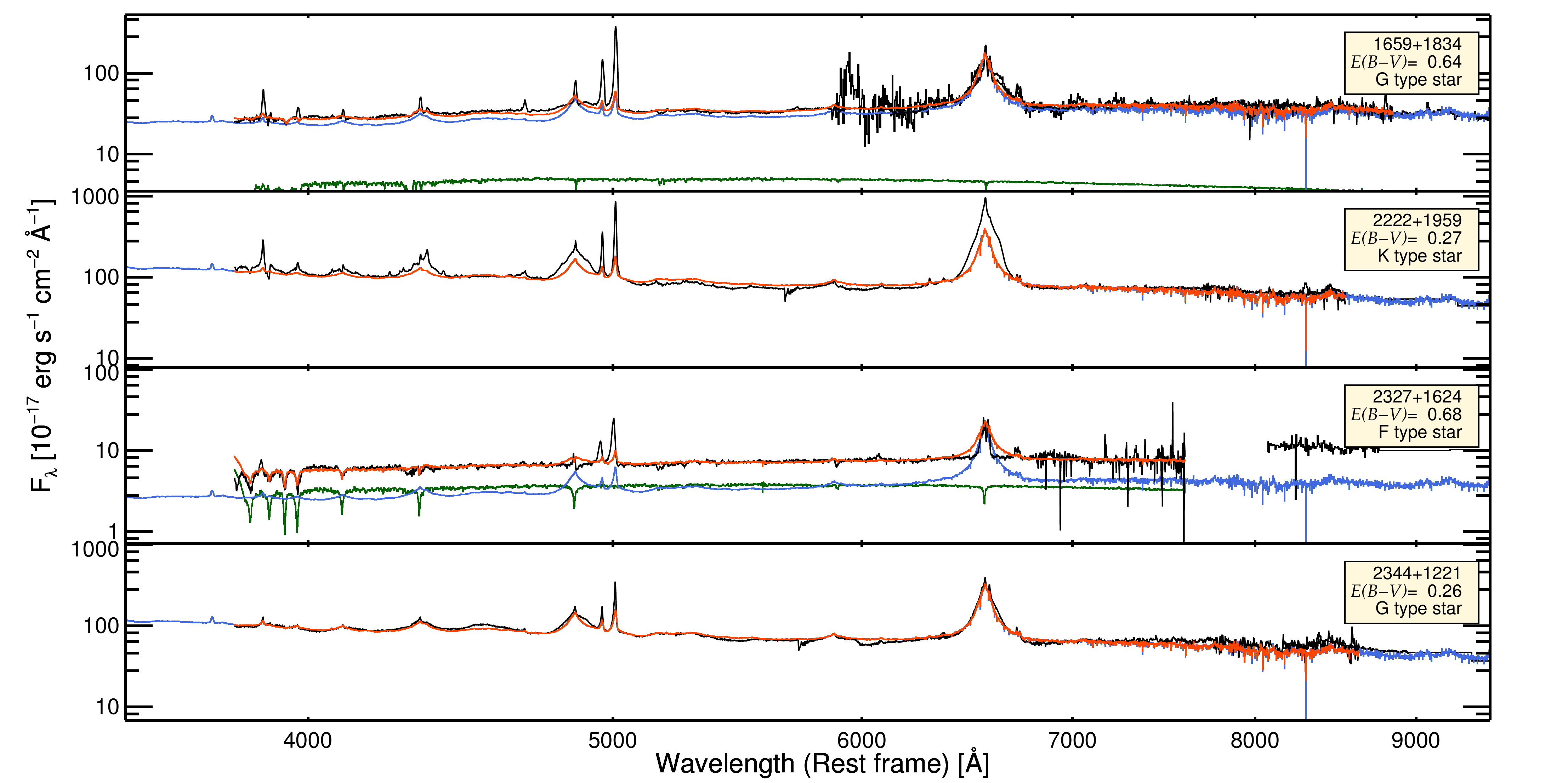}\\
	\caption{Spectra of NIR-red AGNs with the best-fit models are shown
		in the rest-frame from 3790 to 10000\,$\rm \AA{}$.
		The black lines denote the observed spectra.
		The green and blue lines are the best-fit model stellar and quasar spectra, respectively,
		that also include the dust-reddening.
		The red line shows the sum of the best-fit stellar and quasar spectra.
		The top right box denotes the name of NIR-red AGN, the measured $E(B-V)_{\rm cont}$, and the used stellar template.
		Note that the host galaxy dominates the spectra of 0157$+$1712, 0221$+$1327, and 0348$+$1255
		at short wavelength range ($\rm < 6500\,\AA{}$)
		due to heavy extinction of the AGN component.}
\end{figure*}

\section{Accretion rates}
 In this section, we measure the $\lambda_{\rm Edd}$
 ($L_{\rm bol}$/$L_{\rm Edd}$, where $L_{\rm Edd}$ is the Eddington luminosity) of NIR-red AGNs at $z \sim 0.3$.
 To obtain the quantities,
 we derive BH masses and bolometric luminosities
 after correcting for the dust extinction by using the $E(B-V)_{\rm line}$ values
 to avoid the effects of dust extinction.
 If the taken $E(B-V)_{\rm line}$ is negative, we did not correct the dust extinction under
 an assumption that there is no dust extinction.

\begin{figure*}
	\centering
	\figurenum{9}
	\plottwo{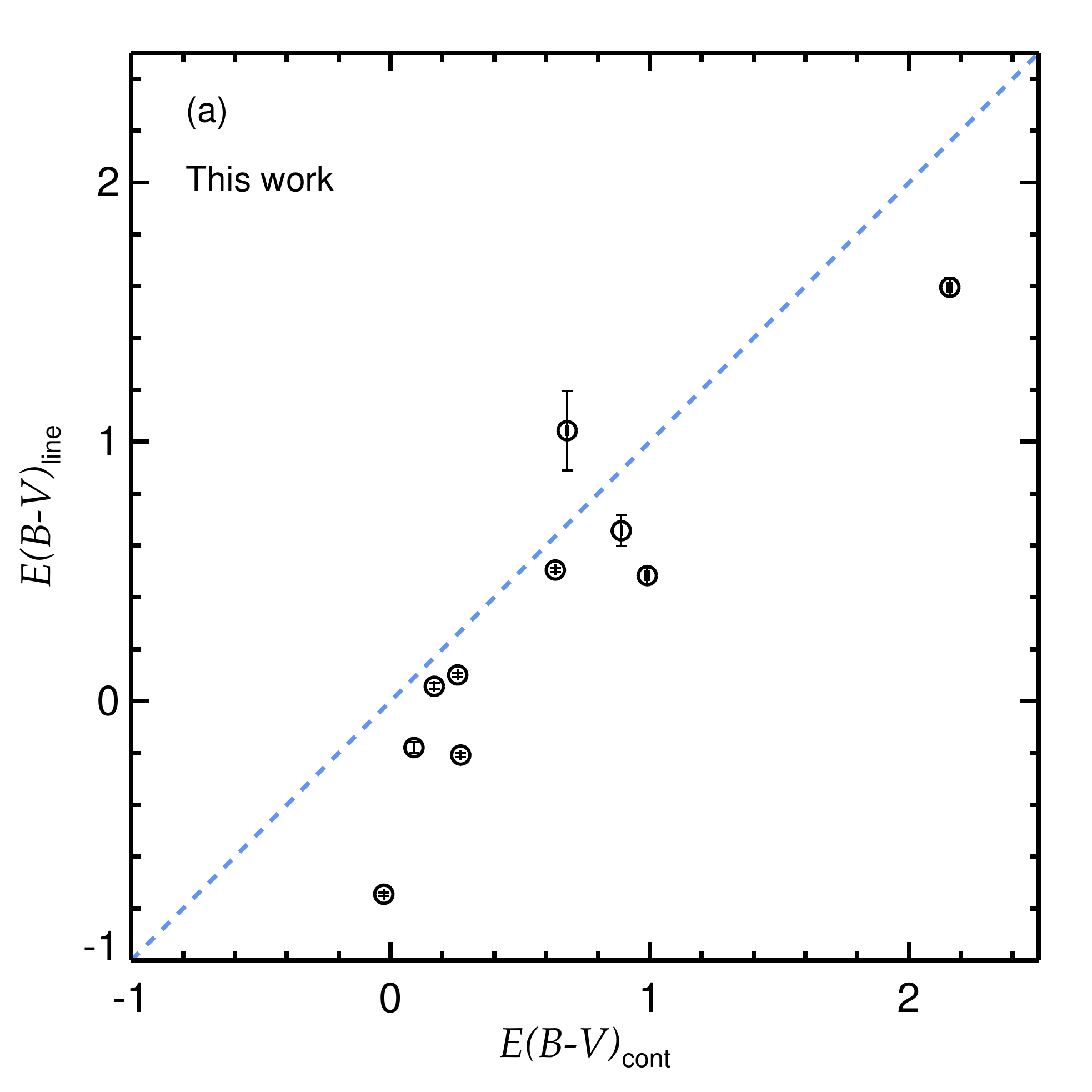}{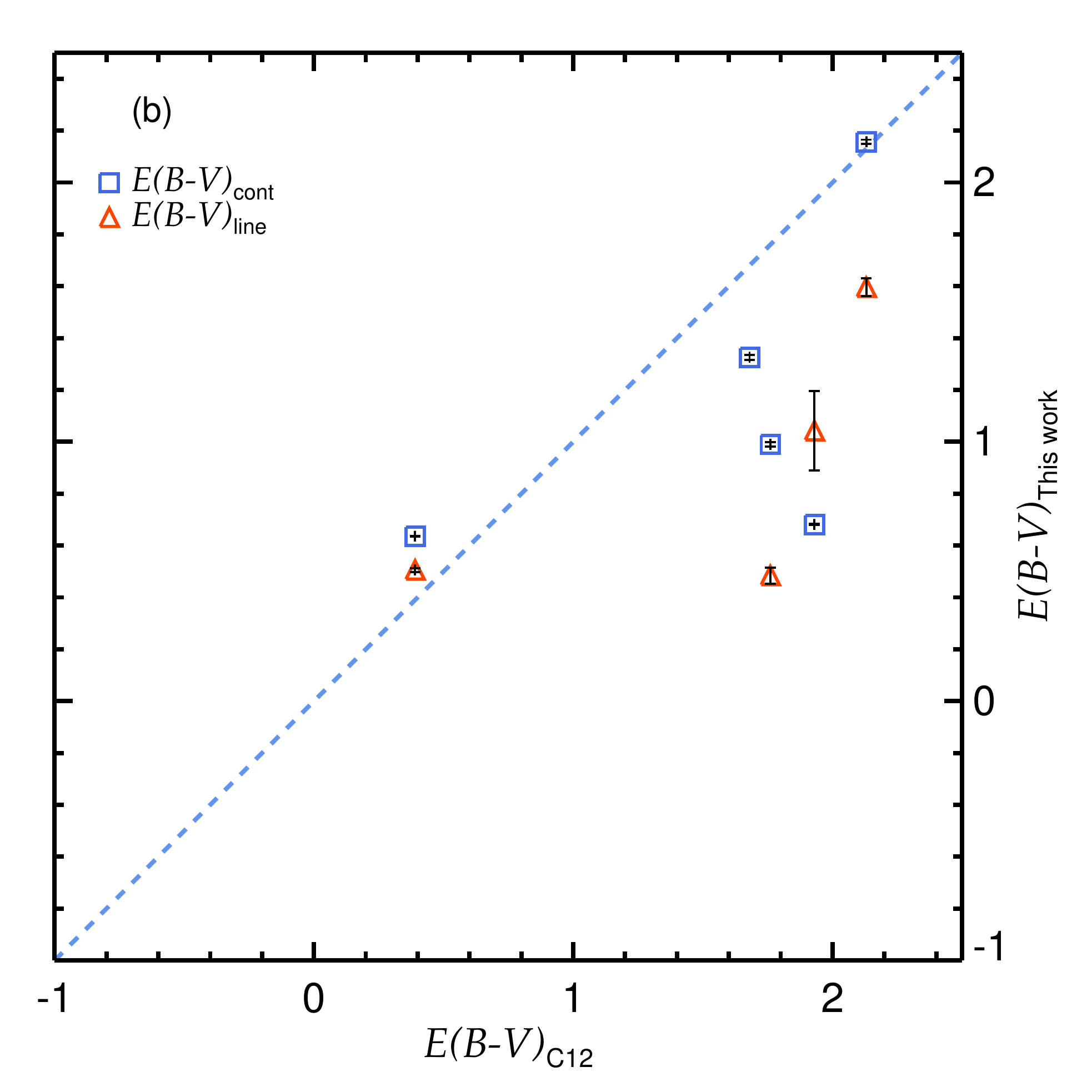}
	\caption{(a) Comparison between the $E(B-V)_{\rm line}$ values and the $E(B-V)_{\rm cont}$ values of NIR-red AGNs.
		The blue dashed line denotes a line where the two values are identical.
		(b) We compare the two types of $E(B-V)$ values with the $E(B-V)_{\rm C12}$ values.
		The blue squares and red triangles represent the $E(B-V)_{\rm cont}$ and $E(B-V)_{\rm line}$ values, respectively.
		The meaning of the blue dashed line is identical to that of the left panel.}
\end{figure*}

 As a comparison sample, we use the unobscured type 1 quasars in
 the quasar property catalog \citep{shen11} of
 the SDSS Seventh Data Release (DR7; \citealt{abazajian09}).
 To avoid the effects of sample bias,
 we set the sample selection criteria to be identical to those of our NIR-red AGNs:
 (i) $0.139 \leq z \leq 0.411$ and (ii) detection in all three 2MASS bands.
 Finally, we select 4130 unobscured type 1 quasars through the sample selection criteria.

 Considering that the $\lambda_{\rm Edd}$ values may have dependence on the $L_{\rm bol}$ values (e.g., \citealt{lusso12,suh15}),
 these selected control samples may cause the sample bias effects.
 Thus, we address this issue in Section 5.3
 by placing restraints on these samples with limited ranges of the $L_{\rm bol}$ and $M_{\rm BH}$.

\subsection{BH masses}

\begin{figure*}
	\centering
	\figurenum{10}
	\includegraphics[scale=0.5]{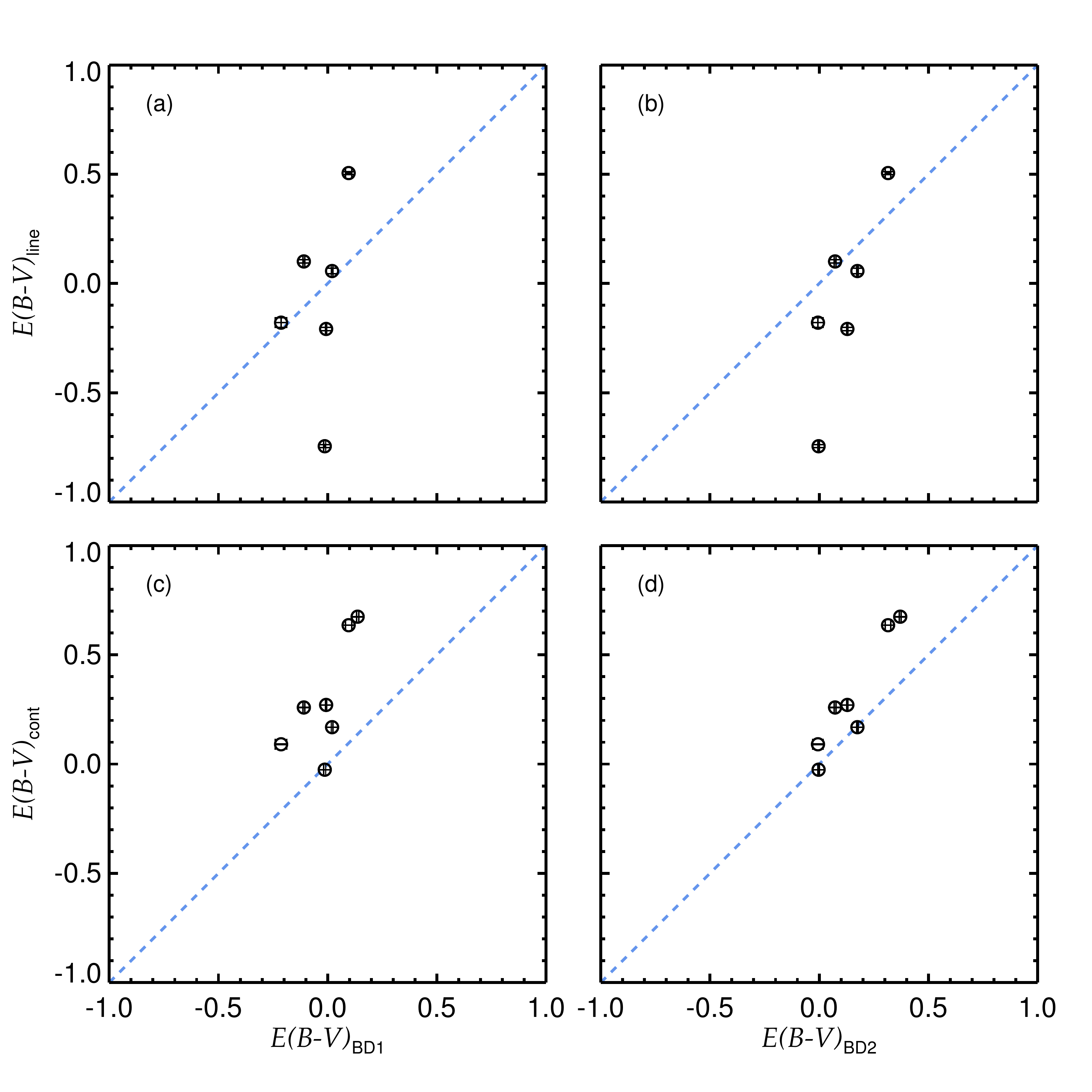}\\
	\caption{(a) Comparison between the $E(B-V)_{\rm BD1}$ and the $E(B-V)_{\rm line}$ values of NIR-red AGNs.
		The meaning of the blue dashed line is identical to Figure 9.
		(b) The $E(B-V)_{\rm BD2}$ vs. $E(B-V)_{\rm line}$ values.
		(c) The $E(B-V)_{\rm BD1}$ vs. $E(B-V)_{\rm cont}$ values.
		(d) The $E(B-V)_{\rm BD2}$ vs. $E(B-V)_{\rm cont}$ values.}
\end{figure*}

\begin{figure*}
	\centering
	\figurenum{11}
	\includegraphics[width=\textwidth]{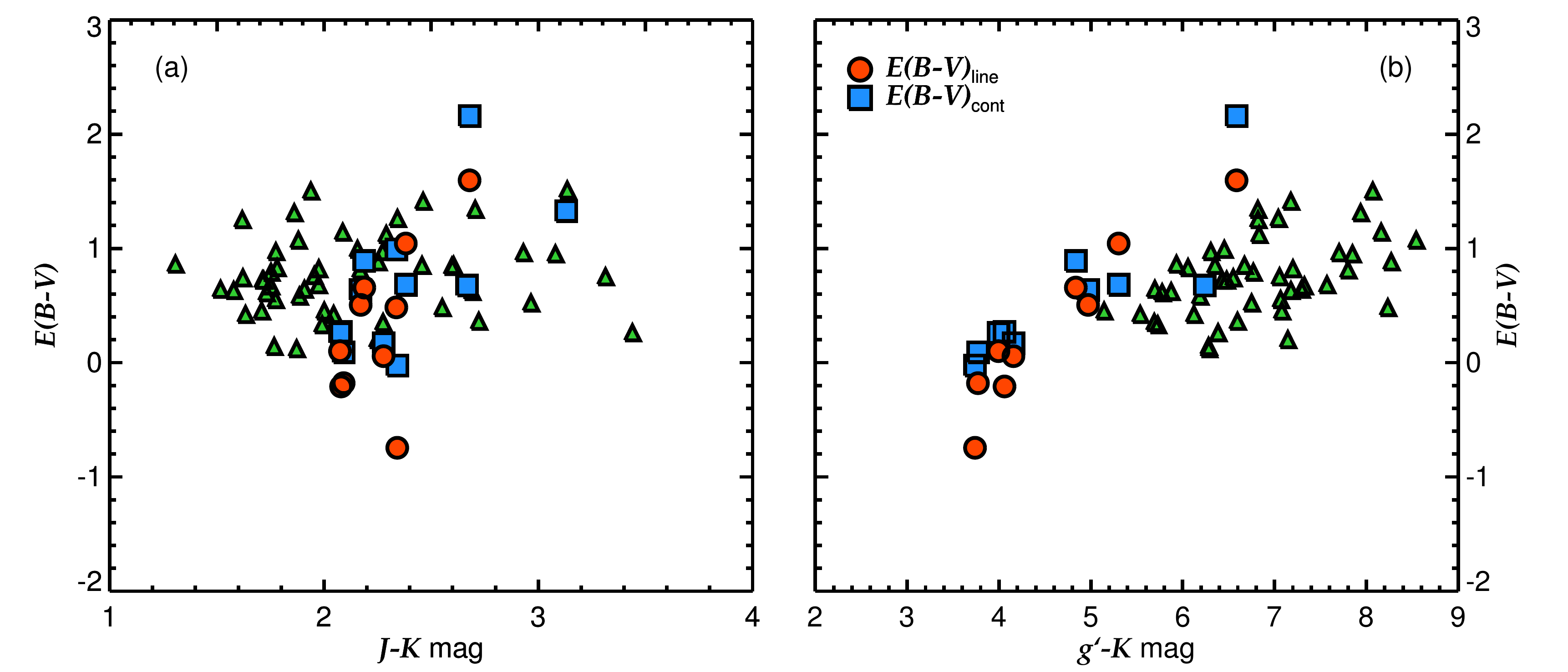}\\
	\caption{(a) Comparison between the measured $E(B-V)$ values and $J-K$ magnitudes.
		The blue squares and red circles represent that the $E(B-V)$ values are the $E(B-V)_{\rm cont}$ and $E(B-V)_{\rm line}$ values, respectively.
		The green triangles show the $J-K$ and $E(B-V)$ values of the red quasars in \cite{urrutia09}.
		(b) Comparison of $g'-K$ magnitudes vs. the $E(B-V)$ values.
		The meanings of the blue squares, red circles, and green triangles are identical to those of the left panel.}
\end{figure*}

 In order to measure the BH masses of NIR-red AGNs,
 NIR $M_{\rm BH}$ estimators (e.g., \citealt{kim10,kim15a,landt11b}) are used
 to alleviate the effects of the dust extinction.
 We adopt the Paschen-line-based $M_{\rm BH}$ estimators \citep{kim10,kim15b},
 to which we applied a recent virial coefficient of $\log f = 0.05$ \citep{woo15}.
 Note that the virial factor, $f$, is the proportional coefficient that is needed to determine the BH mass based on the virial theorem:
\begin{equation}
 M_{\rm BH}=f \frac{R {\Delta V}^{2}}{G},
\end{equation}
 where $R$ is the BLR size, $\Delta V$ is the velocity of the BLR gas, and $G$ is the gravitational constant.
 The value of $\log f$ = 0.05 \citep{woo15} is for the case of $\Delta V$ being the FWHM of the line.
 If the line dispersion is used for $\Delta V$, the $\log f$ value is different.
 The modified, new virial-coefficient-applied, Paschen-line-based $M_{\rm BH}$ estimators are
\begin{equation}
\frac{M_{\rm BH}}{M_{\rm \odot}}=10^{7.04\pm0.02} \left( \frac{L_{\rm P\beta}}{{\rm 10^{42}\,erg\,s^{-1}}} \right)^{0.48\pm0.03}
 \left( \frac{{\rm FWHM_{P\beta}}}{{\rm 10^{3}\,km\,s^{-1}}} \right)^{2}
\end{equation}
 and
\begin{equation}
\frac{M_{\rm BH}}{M_{\rm \odot}}=10^{7.07\pm0.04} \left( \frac{L_{\rm P\alpha}}{{\rm 10^{42}\,erg\,s^{-1}}} \right)^{0.49\pm0.06}
 \left( \frac{{\rm FWHM_{P\alpha}}}{{\rm 10^{3}\,km\,s^{-1}}} \right)^{2}.
\end{equation}

 We measure the BH masses for 11 and 6 NIR-red AGNs by using the P$\beta$- and P$\alpha$-based $M_{\rm BH}$ estimators, respectively.
 There is no significant difference between the measured P$\beta$- and P$\alpha$-based BH masses as shown in Figure 12,
 and the measured BH masses are listed in Table 8.

\begin{figure*}
	\centering
	\figurenum{12}
	\includegraphics[width=\textwidth]{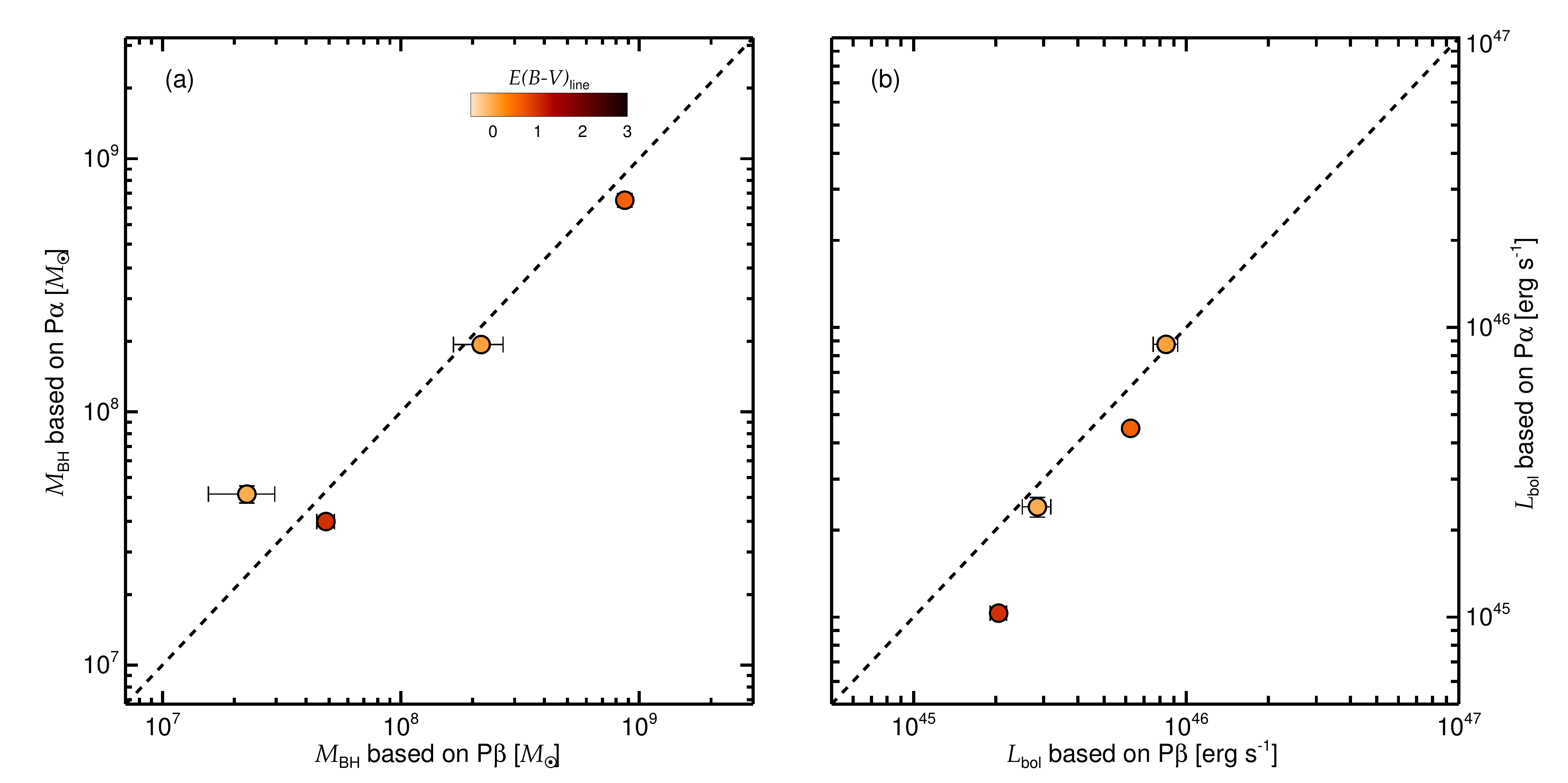}\\
	\caption{(a) Comparison between the measured $M_{\rm BH}$ based on P$\beta$ and P$\alpha$ lines.
		The BH masses are represented by circles, and the colors in the circles mean their $E(B-V)_{\rm line}$ values.
		(b) Comparison of the $L_{\rm bol}$ values derived from P$\beta$ vs. P$\alpha$ lines.
		The circles denote the $L_{\rm bol}$ values, and the meaning of the colors in the circles is identical to the left panel.}
\end{figure*}

 To obtain the BH masses of unobscured type 1 quasars,
 we use an optical $M_{\rm BH}$ estimator \citep{mclure04} consisting of
 $\lambda L_{\rm 5100\,\AA{}}$ (L5100) and $\rm FWHM_{H\beta}$.
 For the optical $M_{\rm BH}$ estimator,
 we apply the virial coefficient of \cite{woo15} as
\begin{equation}
\frac{M_{\rm BH}}{M_{\rm \odot}}=5.27 \left( \frac{{\rm L5100}}{{\rm 10^{44}\,erg\,s^{-1}}} \right)^{0.61}
 \left( \frac{{\rm FWHM_{H\beta}}}{{\rm km\,s^{-1}}} \right)^{2}.
\end{equation}
 The L5100 and $\rm FWHM_{H\beta}$ values of unobscured type 1 quasars are adopted from \cite{shen11}.

\subsection{Bolometric luminosities}

 To estimate the bolometric luminosities of NIR-red AGNs,
 we combine several empirical relations between the bolometric luminosity ($L_{\rm bol}$),
 the continuum luminosity, and the line luminosity.
 We bootstrap the empirical relations between
 $L_{\rm bol}$ and L5100 \citep{shen11},
 L5100 and $L_{\rm H\alpha}$ \citep{jun15},
 and $L_{\rm H\alpha}$ and the two Paschen line luminosities \citep{kim10}.
 The combined relations between $L_{\rm bol}$ and the Paschen line luminosities are
\begin{equation}
\log \left( \frac{L_{\rm bol}}{\rm{10^{44}\,erg\,s^{-1}}} \right)=1.33+0.966\,\log \left( \frac{L_{\rm P\beta}}{\rm{10^{42}\,erg\,s^{-1}}} \right)
\end{equation}
and
\begin{equation}
\log \left( \frac{L_{\rm bol}}{\rm{10^{44}\,erg\,s^{-1}}} \right)=1.27+0.920\,\log \left( \frac{L_{\rm P\alpha}}{\rm{10^{42}\,erg\,s^{-1}}} \right).
\end{equation}

 We measured the $L_{\rm bol}$ values for 12 and 6 NIR-red AGNs by using $L_{\rm P\beta}$ and $L_{\rm P\alpha}$, respectively.
 The $L_{\rm bol}$ measured from P$\beta$ and P$\alpha$ show no significant differences, as shown in Figure 12,
 and the measured $L_{\rm bol}$ values are listed in Table 8.

 To obtain the $L_{\rm bol}$ values of unobscured type 1 quasars,
 we use L5100 values, with the bolometric correction factor (9.26) for L5100,
 both of which are adopted from \cite{shen11}.

\subsection{Eddington ratios of NIR-red AGNs}

\begin{figure*}
	\centering
	\figurenum{13}
	\includegraphics[scale=0.5]{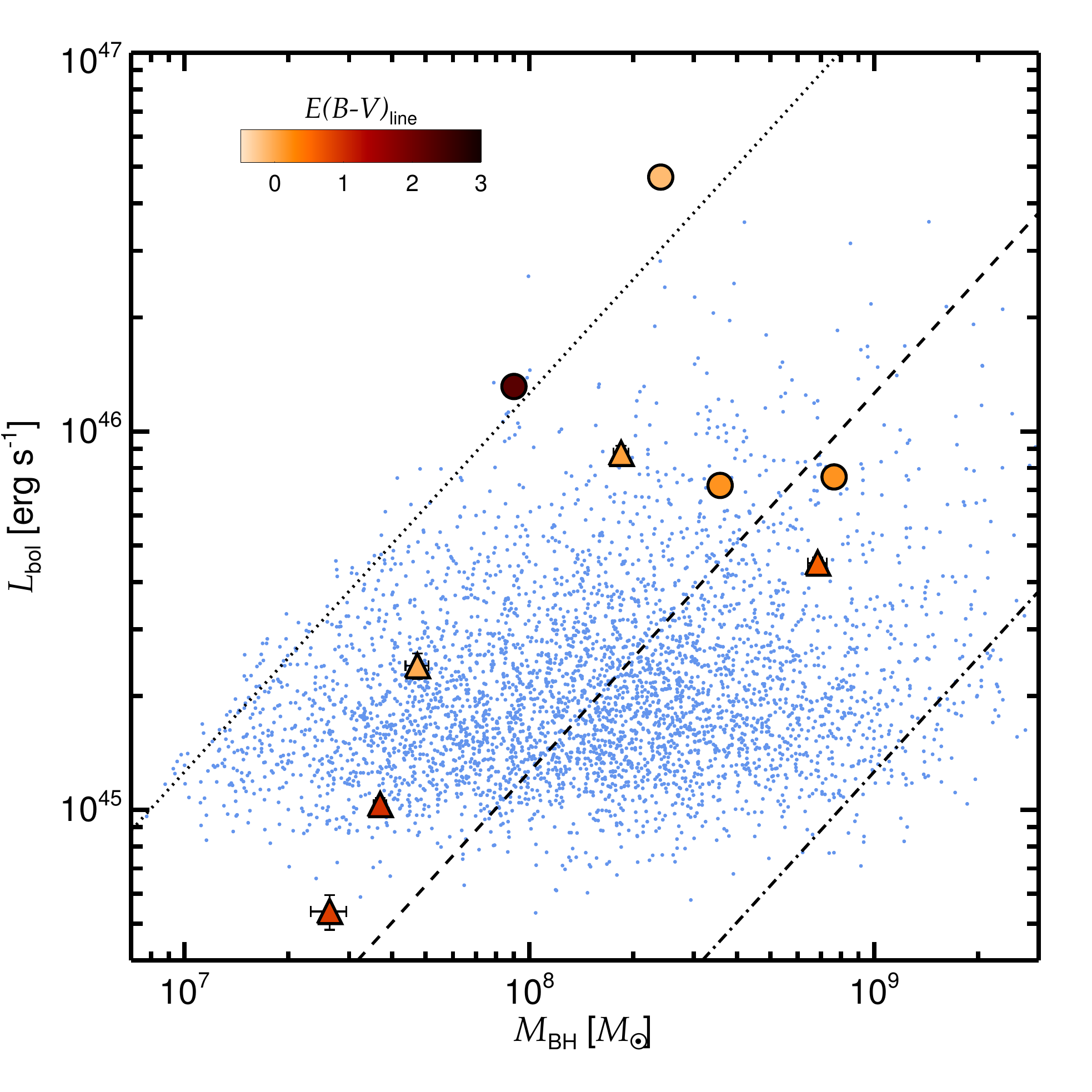}\\
	\caption{The $L_{\rm bol}$ and $M_{\rm BH}$ values of NIR-red AGNs and unobscured type 1 quasars.
		The circles and triangles show the $L_{\rm bol}$ and $M_{\rm BH}$ values of NIR-red AGNs
		derived from P$\beta$ and P$\alpha$, respectively.
		The meaning of the colors in the circles and triangles is identical to that in Figure 12.
		The blue dots represent the $L_{\rm bol}$ and $M_{\rm BH}$ values of unobscured type 1 quasars.
		The dotted, dashed, and dash-dotted lines denote $\lambda_{\rm Edd}$ of 1, 0.1, and 0.01, respectively.}
\end{figure*}

\begin{figure}
	\centering
	\figurenum{14}
	\includegraphics[scale=0.35]{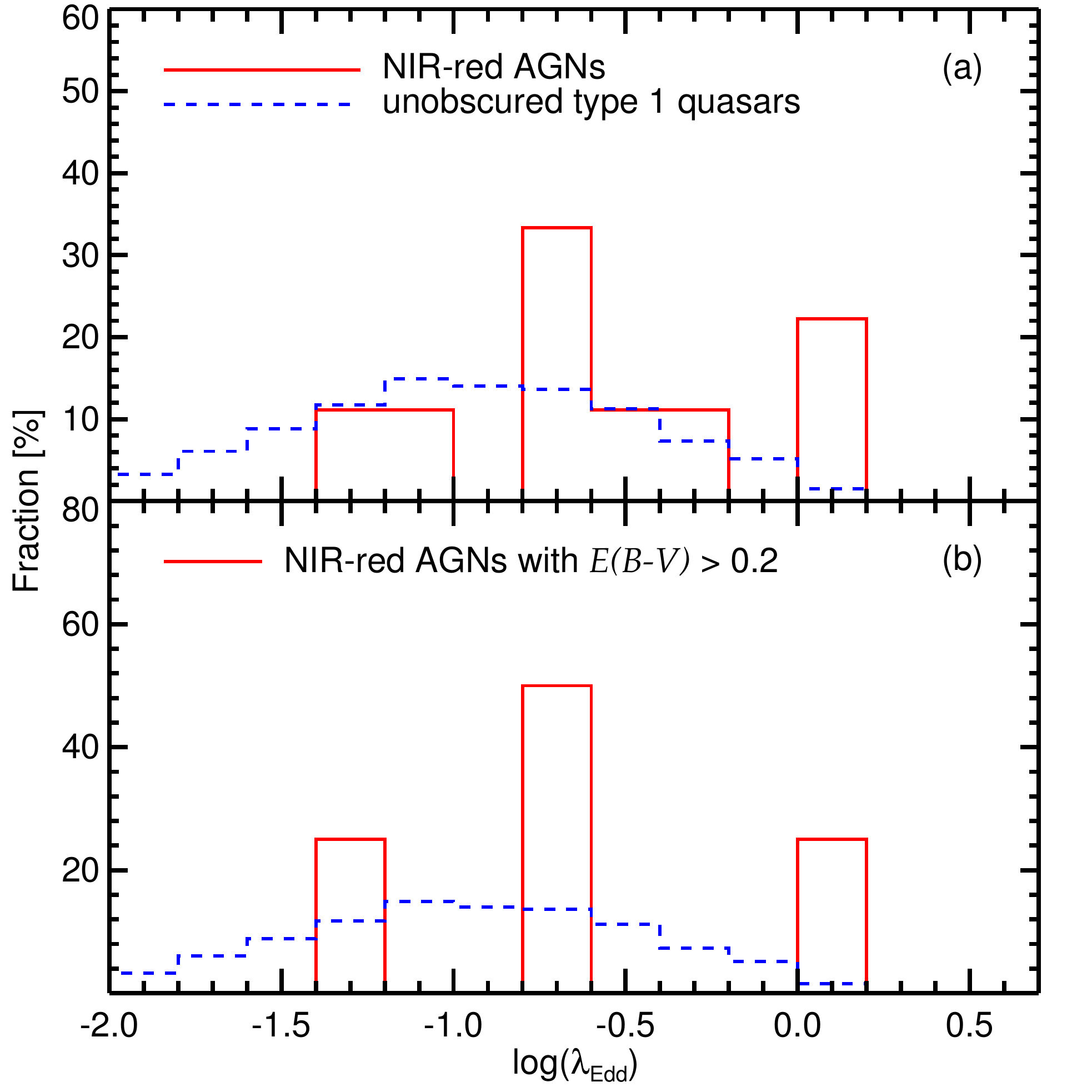}\\
	\caption{(a) Distributions of the $\lambda_{\rm Edd}$ values of NIR-red AGNs and unobscured type 1 quasars.
		The red solid and blue dashed histograms show the $\lambda_{\rm Edd}$ distributions of NIR-red AGNs and unobscured type 1 quasars, respectively.
		(b) Distributions of the $\lambda_{\rm Edd}$ of NIR-red AGNs with $E(B-V) > 0.2$ and unobscured type 1 quasars.
		The meaning of the red solid lines, blue dashed lines, and the ranges of the $L_{\rm bol}$ and $M_{\rm BH}$
		are identical to those in panel (a).}
\end{figure}

 When comparing the $\lambda_{\rm Edd}$ values of NIR-red AGNs to those of unobscured type 1 quasars,
 we prefer to use the $\lambda_{\rm Edd}$ values from P$\alpha$ than those from P$\beta$
 when both quantities are available, to minimize the effects from the dust extinction.
 The $\lambda_{\rm Edd}$ values of nine NIR-red AGNs are used for the comparison,
 among which four (0157$+$1712, 0324$+$1748, 2222$+$1959, and 2344$+$1221) and
 five (0221$+$1327, 1258$+$2329, 1453$+$1353, 1543$+$1937, and 1659$+$1834)
 $\lambda_{\rm Edd}$ values are derived from P$\beta$ and P$\alpha$, respectively.

 The $M_{\rm BH}$ and $L_{\rm bol}$ values of NIR-red AGNs and unobscured type 1 quasars are shown in Figure 13,
 and Figure 14 shows their distributions of $\lambda_{\rm Edd}$ values.
 We find that the median $\lambda_{\rm Edd}$ of the nine NIR-red AGNs, $\log (\lambda_{\rm Edd}) = -0.654 \pm 0.174$,
 where the error represents the error of the median,
 is only mildly higher than that of unobscured type 1 quasars, $\log (\lambda_{\rm Edd}) = -0.961 \pm 0.008$.
 For quantifying how significantly these two distributions of the $\lambda_{\rm Edd}$ values differ from each other,
 we perform the Kolmogorov–Smirnov test (K-S test) by using the \texttt{KSTWO} code based on IDL.
 The maximum deviation between the cumulative distributions of these two $\lambda_{\rm Edd}$ values, $D$, is 0.392,
 and the probability of the result given the null hypothesis, $p$, is 0.094.

 From this comparison, we conclude that the $\lambda_{\rm Edd}$ of the NIR-red AGNs is only slightly larger than
 that of the unobscured type 1 AGNs, but statistically the difference is not significant.
 A few outliers with large $\lambda_{\rm Edd}$ appear to dominate the K-S test result,
 and this suggests that the NIR-red AGN sample is mostly indiscernible in their property from unobscured type 1 AGNs,
 but can include truly dusty high $\lambda_{\rm Edd}$ AGNs such as 0157$+$1712 with $E(B-V)_{\rm line} = 1.596$.

 Since $\lambda_{\rm Edd}$ values can depend on the $L_{\rm bol}$ values (e.g., \citealt{lusso12,suh15}),
 we repeated the analysis after matching their $L_{\rm bol}$ values.
 First, we divide the NIR-red AGNs into two sub-samples, four low-$L_{\rm bol}$
 (0221$+$1327, 1258$+$2329, 1453$+$1353, and 1659$+$1834; $44.73 \leq \log( L_{\rm bol} / {\rm erg\,s^{-1}}) \leq 45.65$) and
 five high-$L_{\rm bol}$ (0157$+$1712, 0324$+$1748, 1543$+$1937, 2222$+$1959, and 2344$+$1221;
 ${45.86} \leq \log( L_{\rm bol} / {\rm erg\,s^{-1}}) \leq 46.67$) NIR-red AGNs.
 Second, among all the unobscured type 1 AGNs,
 we choose 3688 low-$L_{\rm bol}$ ($44.76 \leq \log( L_{\rm bol} / {\rm erg\,s^{-1}}) \leq 45.65$)
 and 165 high-$L_{\rm bol}$ (${45.86} \leq \log( L_{\rm bol} / {\rm erg\,s^{-1}}) \leq 46.55$) sub-samples
 that have the similar $L_{\rm bol}$ ranges to those of the divided NIR-red AGNs.

 By comparing such the $L_{\rm bol}$ limited samples, we confirm the result from the full sample,
 which is the $\lambda_{\rm Edd}$ of NIR-red AGNs is only mildly higher than that of unobscured type 1 AGNs.
 For the low-$L_{\rm bol}$ samples, the median $\log(\lambda_{\rm Edd})$ values of the NIR-red AGNs and unobscured type 1 AGNs are
 $-0.654 \pm 0.216$ and $-0.995 \pm 0.008$, respectively.
 For the high-$L_{\rm bol}$ samples, although the $\lambda_{\rm Edd}$ values are larger than those of the low-$L_{\rm bol}$ samples,
 the result is consistent throughout.
 The median $\log(\lambda_{\rm Edd})$ of the NIR-red AGNs is $-0.424 \pm 0.276$,
 and that of the unobscured type 1 AGNs is ${-0.661} \pm {0.035}$.

 We also compare the $\lambda_{\rm Edd}$ of unobscured type 1 quasars to those of NIR-red AGNs with $E(B-V) > 0.2$.
 The comparison is shown in Figure 14.
 We find that the median $\log(\lambda_{\rm Edd})$ of the NIR-red AGNs is -0.654$\pm$0.321,
 which is consistent with the above results.

\section{$M_{\rm BH}$--$\sigma_{\ast}$ relation}

\begin{figure*}
	\centering
	\figurenum{15}
	\includegraphics[scale=0.5]{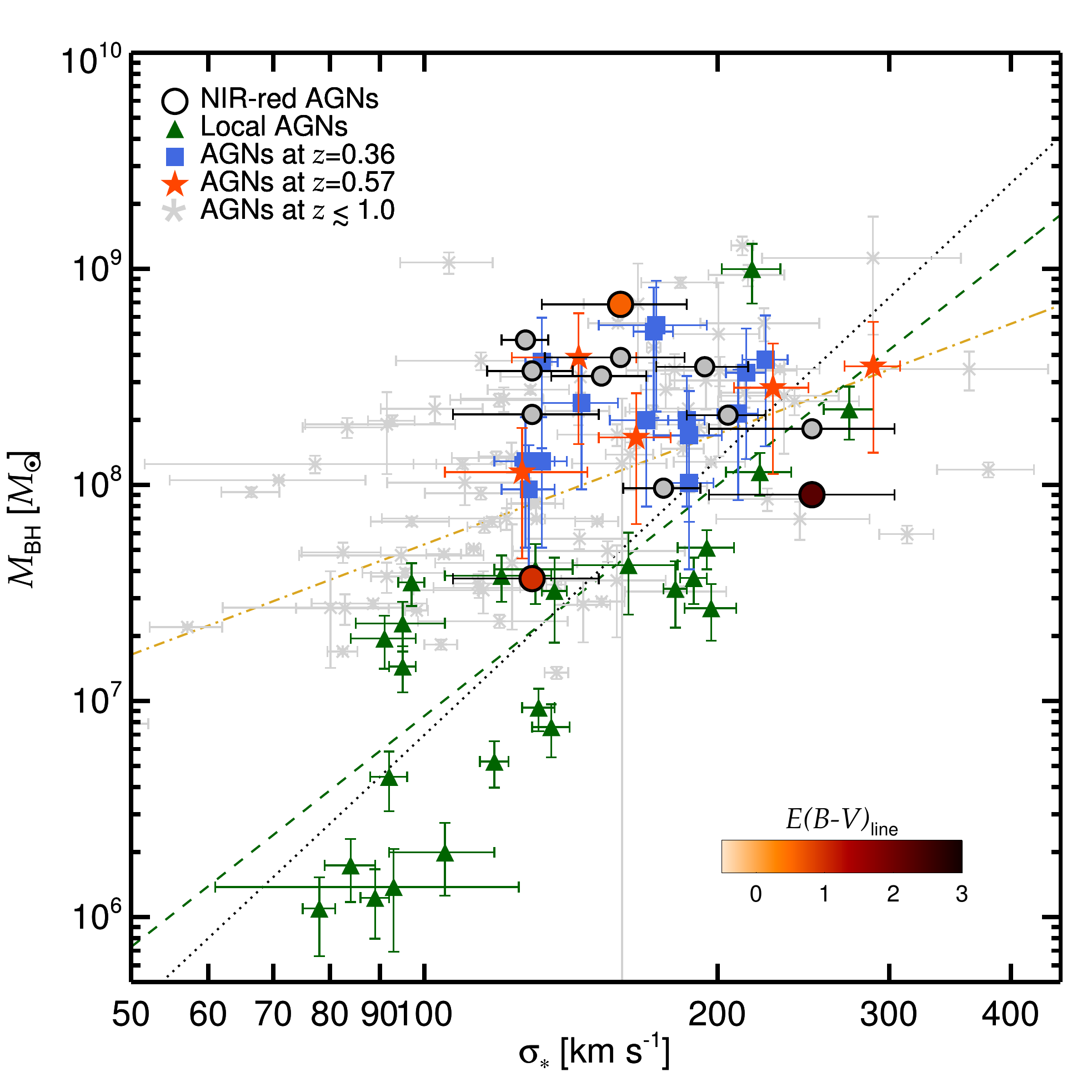}\\
	\caption{$M_{\rm BH}$--$\sigma_{\ast}$ relations of AGNs.
		Circles denote the $M_{\rm BH}$ and $\sigma_{\ast}$ values of NIR-red AGNs,
		and the meaning of the colors in the circles is identical to Figure 11.
		The gray circles denote the quantities of NIR-red AGNs measured in \cite{canalizo12}.
		The $M_{\rm BH}$ and $\sigma_{\ast}$ values of unobscured type 1 AGNs at
		$z \simeq$ 0 \citep{woo10}, 0.36 \citep{woo06}, 0.57 \citep{woo08}, and $\lesssim$1 \citep{shen15}
		are represented by green triangles, blue squares, red stars, and gray asterisks, respectively.
		The green dashed, black dotted, and yellow dot-dashed lines show
		the $M_{\rm BH}$--$\sigma_{\ast}$ relations for local AGNs \citep{woo10}, quiescent local galaxies \citep{gultekin09},
		and unobscured type 1 AGNs at $z \sim 0.26$ \citep{shen15}.}
\end{figure*}

\begin{figure*}
	\centering
	\figurenum{16}
	\includegraphics[scale=0.4]{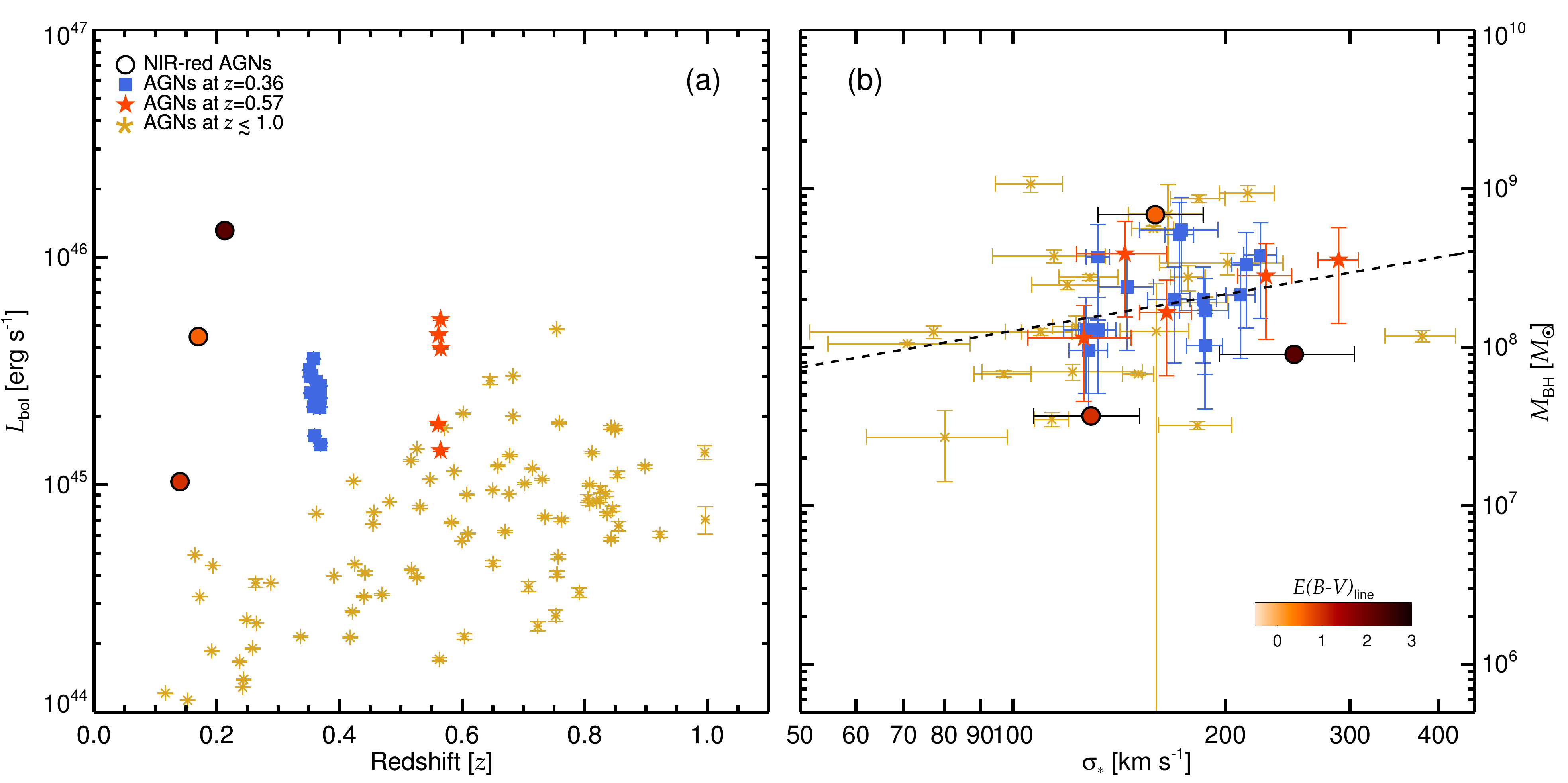}\\
	\caption{
		(a) The $L_{\rm bol}$ values vs. redshifts of NIR-red AGNs and unobscured type 1 AGNs.
		The circles denote the redshifts and $L_{\rm bol}$ values of NIR-red AGNs,
		and the meaning of the colors in the circles is identical to those in Figure 11.
		The blue squares, red stars, and yellow asterisks represent the quantities of unobscured type 1 AGNs
		at $z \simeq$ 0.36 \citep{woo06}, 0.57 \citep{woo08}, and $\lesssim$1 \citep{shen15}, respectively.
		(b) $M_{\rm BH}$--$\sigma_{\ast}$ relations of the luminosity-matched AGNs.
		The meaning of circles, blue squares, red stars, and yellow asterisks have meanings identical to those in the left panel.
		The black dashed line indicates the $M_{\rm BH}$--$\sigma_{\ast}$ relation of the luminosity-matched unobscured type 1 AGNs,
		where the measured $\alpha$ and $\beta$ are 8.335$\pm$0.016 and 0.768$\pm$0.083, respectively.
	}
\end{figure*}

 In this section, we discuss the $M_{\rm BH}$--$\sigma_{\ast}$ relation of NIR-red AGNs.
 For the relation, we adopt the $\sigma_{\ast}$ values from \cite{canalizo12} which are measured
 using the stellar absorption lines in the range of 3900--5500\,$\rm \AA{}$.
 However, the $\sigma_{\ast}$ values are available for only three objects (0157$+$1712, 0221+1327, and 1659+1834) in our sample.
 These three NIR-red AGNs have $E(B-V) > 0.2$.

 For the BH masses, \cite{canalizo12} also estimated the BH masses using the L5100 and $\rm FWHM_{\rm H\alpha}$ values,
 but we use the Paschen-line-based BH masses measured in Section 5.2.
 We compare these two kinds of BH masses.
 In \cite{canalizo12}, the BH masses based on L5100 and $\rm FWHM_{\rm H\alpha}$
 are $10^{8.35}$, $10^{8.42}$, and $10^{8.68}$\,$M_{\rm \odot}$ for 0157$+$1712, 0221+1327, and 1659+1834, respectively,
 whereas the BH masses derived from the Paschen lines are ${10^{7.96}}$, $10^{7.57}$, and $10^{8.84}$\,$M_{\rm \odot}$.
 Although there is no significant difference in the BH masses for 0157$+$1712 and 1659+1834,
 the Paschen-line-based BH mass of 0221+1327 is smaller by a factor of $\sim 7$.
 The discrepancy for the BH mass of 0221$+$1327 does not come from the dust extinction but from the spectral line fitting.
 In \cite{canalizo12}, they measured the FWHM as 4279\,$\rm km\,s^{-1}$, which is estimated from the H$\alpha$ line.
 In this study, we measured the FWHM from the H$\alpha$, P$\beta$, and P$\alpha$ lines,
 which gives 2212, 2125, and 2073\,$\rm km\,s^{-1}$, respectively,
 and these values are significantly smaller than the previous result.

 In Figure 15, the newly established $M_{\rm BH}$--$\sigma_{\ast}$ relation of NIR-red AGNs
 is presented, along with those for local quiescent galaxies \citep{gultekin09} and
 unobscured type 1 AGNs at $z \simeq$ 0, 0.36, 0.57, and $\lesssim 1$ \citep{woo06,woo08,woo10,shen15}.
 These $M_{\rm BH}$--$\sigma_{\ast}$ relations of unobscured type 1 AGNs are modified
 by applying the virial coefficient of $\log f = 0.05$ \citep{woo15}, as we did for NIR-red AGNs.

 By comparing the $M_{\rm BH}$--$\sigma_{\ast}$ relation of NIR-red AGNs and local unobscured type 1 AGNs \citep{woo10},
 we find offsets of $\Delta \log (M_{\rm BH}/M_{\rm \odot}) =$
 -0.389, 0.243, and 1.190 for 0157$+$1712, 0221+1327, and 1659+1834, respectively,
 resulting in a mean offset of $\Delta \log (M_{\rm BH}/M_{\rm \odot}) = {0.348 \pm 0.902}$. 
 The $M_{\rm BH}$--$\sigma_{\ast}$ relation for NIR-red AGNs and those for unobscured type 1 AGNs at $z = 0$ through 0.5
 are consistent with each other
 (e.g., $\Delta \log (M_{\rm BH}/M_{\rm \odot}) = $ 0.74, 0.62, and 0.52 for unobscured type 1 AGNs
 at $z=$ 0.26, 0.36, and 0.57, respectively; \citealt{shen15,woo06,woo08}).
 This result suggests that there is no significant offset in the $M_{\rm BH}$--$\sigma_{\ast}$ relation between
 the NIR-red AGNs and the unobscured type 1 AGNs,
 although more objects are needed to better quantify the offset.

 Moreover, we compare the $M_{\rm BH}$--$\sigma_{\ast}$ relations of NIR-red AGNs and unobscured type 1 AGNs
 after matching their $L_{\rm bol}$ values to exclude the selection bias introduced from their different luminosities (e.g., \citealt{shen15}).
 The $L_{\rm bol}$ values of 0157$+$1712, 0221$+$1327, and 1659$+$1834 are
 $10^{46.12}$, $10^{45.01}$, and $10^{45.65}$\,$\rm erg\,s^{-1}$, respectively, as shown in Figure 16,
 and the $L_{\rm bol}$ values of the unobscured type 1 AGNs at $z \simeq$ 0.36 \citep{woo06}, 0.57 \citep{woo08}, and $\lesssim 1$ \citep{shen15}
 are measured by applying the bolometric correction factor of 9.26 \citep{shen11} to their L5100 values.
 Among them, we choose 42 unobscured type 1 AGNs that have similar, but somewhat lower, $L_{\rm bol}$ range
 ($10^{45.01} \leq L_{\rm bol}/{\rm erg\,s^{-1}} \leq 10^{45.73}$), to that of the NIR-red AGNs.
 For the selected unobscured type 1 AGNs, their $M_{\rm BH}$ can be expressed as a function of $\sigma_{\rm \ast}$:
\begin{equation}
\log \left( \frac{M_{\rm BH}}{M_{\rm \odot}} \right) = \alpha + \beta \,\log \left( \frac{\sigma_{\rm \ast}}{{\rm 200\,km\,s^{-1}}} \right),
\end{equation}
 and the measured $\alpha$ and $\beta$ are 8.335$\pm$0.016 and 0.768$\pm$0.083, respectively.
 By comparing the newly measured $M_{\rm BH}$--$\sigma_{\rm \ast}$ relation of the unobscured type 1 AGNs and that of the NIR-red AGNs,
 we find a mean offset of $\Delta \log (M_{\rm BH}/M_{\rm \odot}) = {-0.166 \pm 0.681}$, as presented in Figure 16.
 In this luminosity-matched comparison, the number of the NIR-red AGNs is also insufficient,
 but we obtain the same result that there is no significant offset between
 the $M_{\rm BH}$--$\sigma_{\ast}$ relations of the NIR-red AGNs and the unobscured type 1 AGNs.

\section{Summary}
 By performing NIR spectroscopic observations with
 four telescopes, Gemini, IRTF, Magellan, and Subaru,
 we obtained 0.7--2.5\,$\mu$m medium-resolution ($R > 2000$) and high-S/N (up to several hundreds) spectra of 16 NIR-red AGNs at $z \sim 0.3$.
 In addition to the NIR spectra,
 we obtained optical (0.4--1.0\,$\mu$m) medium-resolution ($R \sim 4000$) spectra of 12 NIR-red AGNs taken with Keck/ESI and SDSS data.
 Using both sets of spectra, we measured the line luminosities and FWHMs of H$\beta$, H$\alpha$, P$\beta$, and P$\alpha$ lines
 for 7, 12, 12, and 6 NIR-red AGNs, respectively.

\begin{figure*}[!t]
	\centering
	\includegraphics[width=\textwidth]{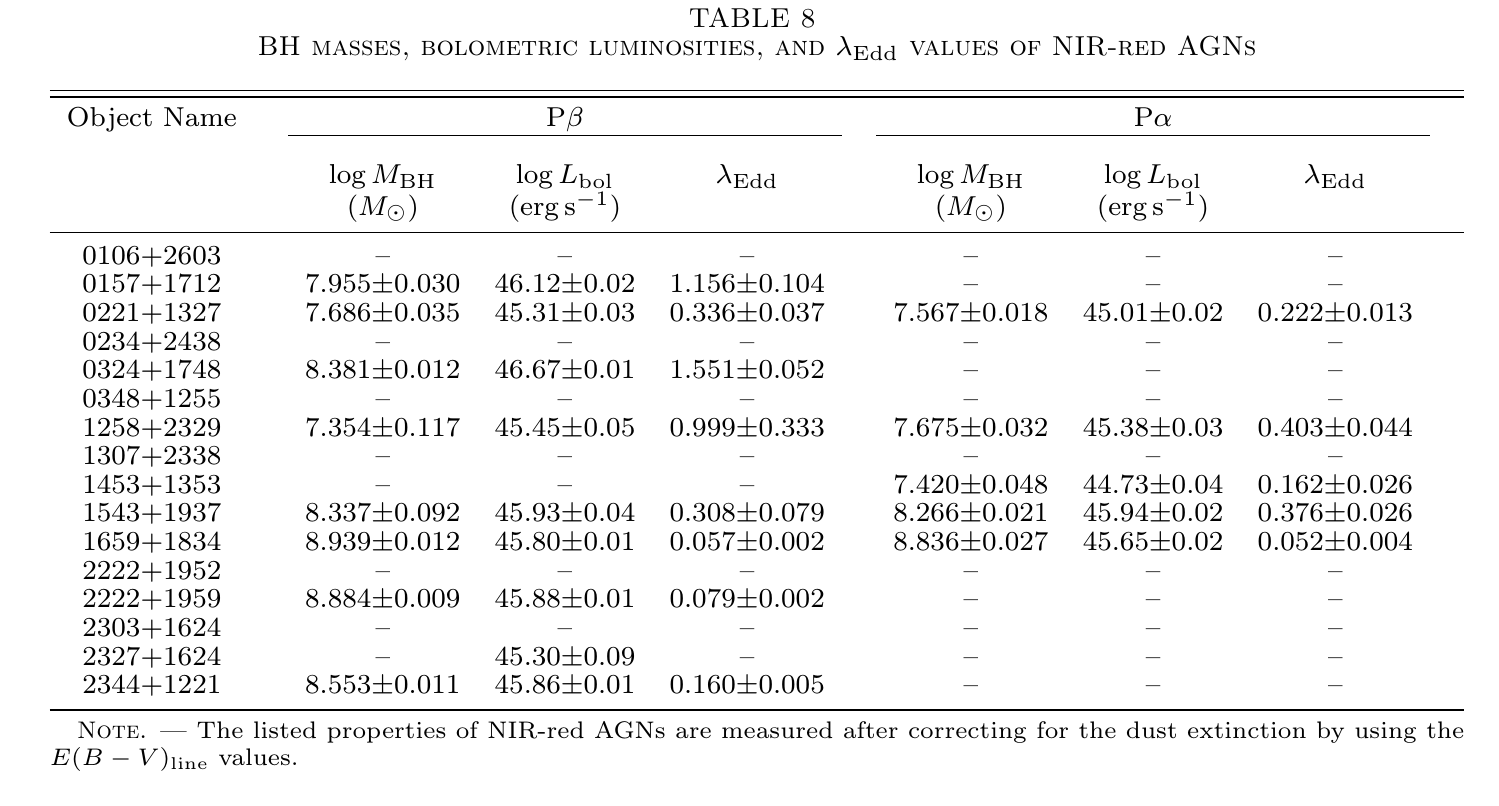}\\
\end{figure*}

 Before analyzing the physical properties of NIR-red AGNs,
 we derived the $E(B-V)$ values of NIR-red AGNs in two ways.
 First, we estimated the $E(B-V)_{\rm line}$ values
 by using the luminosity ratios of H$\beta$, H$\alpha$, P$\beta$, and P$\alpha$ lines.
 Second, the $E(B-V)_{\rm cont}$ values were measured
 by comparing the continuum slopes at 3790--10000\,$\rm \AA{}$.
 Through these two methods, we measured the $E(B-V)_{\rm line}$ and $E(B-V)_{\rm cont}$ values for 10 and 12 NIR-red AGNs, respectively.
 These two $E(B-V)$ values are consistent, and their Pearson correlation coefficient is 0.911.

 Among our sample, five objects have low $E(B-V) < 0.2$.
 Comparing the previous result \citep{urrutia09}
 from the optical-NIR and NIR color selection that yields only two objects have low $E(B-V)$ values in $\sim 50$ candidates,
 we suspect that the NIR red color selection alone is not effective at picking up dusty red AGNs.

 After correcting for the dust extinction with the measured $E(B-V)$ values,
 we measured the $\lambda_{\rm Edd}$ values of NIR-red AGNs.
 For the $M_{\rm BH}$ and $L_{\rm bol}$ values, we used Paschen-line-based $M_{\rm BH}$ and $L_{\rm bol}$ estimators
 to alleviate the effects of the dust extinction.
 The newly estimated median $\lambda_{\rm Edd}$ of NIR-red AGNs, $\log ({\rm Edd}) = -0.654 \pm 0.174$, is only mildly higher than
 that of unobscured type 1 quasars, $\log ({\rm Edd}) = -0.961 \pm 0.008$.

 Using the measured BH masses,
 we compared the $M_{\rm BH}$--$\sigma_{\ast}$ relation of NIR-red AGNs to that of unobscured type 1 AGNs at similar redshift.
 Although only three objects were used,
 NIR-red AGNs show a tendency to have similar BH masses at a fixed the $\sigma_{\ast}$.

 Our results suggest that AGNs with red $J-K$ colors are not necessarily dust-obscured AGNs,
 and the selection of dusty AGNs needs to be carefully performed.

\acknowledgements
 We thank the referee for useful comments.
 This work was supported by the Creative Initiative Program of the National Research Foundation of Korea (NRF),
 No. 2017R1A3A3001362, funded by the Korea government (MSIP).
 D.K. acknowledges support by the National Research Foundation of Korea to
 the Fostering Core Leaders of the Future Basic Science Program, No. 2017-002533.
 The Gemini data were taken through the K-GMT Science Program
 (PID: GN-2015B-Q-51; GN-2016A-Q-86) of Korea Astronomy and Space Science Institute (KASI).
 This paper includes data obtained with the 6.5\,m Magellan Telescopes
 located at Las Campanas Observatory, Chile.
 D.K. and M.I. are Visiting Astronomers (PID: 2015B092; 2016A043) at the Infrared Telescope Facility,
 which is operated by the University of Hawaii under Cooperative Agreement no. NNX-08AE38A
 with the National Aeronautics and Space Administration,
 Science Mission Directorate, Planetary Astronomy Program. 
 This paper includes data obtained with Subaru telescope and W. M. Keck observatory on Maunakea.



\clearpage

\end{document}